# Inverse Spin Hall Effect from pulsed Spin Current in Organic Semiconductors with Tunable Spin-Orbit Coupling

Dali Sun[§], Kipp J. van Schooten[§], Hans Malissa, Marzieh Kavand, Chuang Zhang, Christoph Boehme*, and Z. Valy Vardeny*

*Department of Physics & Astronomy, University of Utah, Salt Lake City, Utah, 84112, USA*

**Exploration of spin-currents in organic semiconductors (OSECs) induced by resonant microwave absorption in ferromagnetic substrates has been of great interest for potential spintronics applications. Due to the inherently weak spin-orbit coupling (SOC) of OSECs, their inverse spin Hall effect (ISHE) response is very subtle; limited by the microwave power applicable under continuous-wave (cw) excitation. Here we introduce a novel approach for generating significant ISHE signals using pulsed ferromagnetic resonance, where the ISHE is ~2-3 orders of magnitude larger compared to cw excitation. This strong ISHE enables us to investigate a variety of OSECs ranging from π-conjugated polymers with strong SOC that contain intrachain platinum atoms, to weak SOC polymers, to $C_{60}$ films, where the SOC is predominantly caused by the molecule surface curvature. The pulsed-ISHE technique offers a robust route for efficient injection and detection schemes of spin-currents at room temperature, and paves the way for spin-orbitronics in plastic materials.**

[§]These authors contributed equally to this work.

*To whom correspondence should be addressed: boehme@physics.utah.edu, val@physics.utah.edu

Coupled charge and spin transport phenomena have drawn great attention in the past few years as they allow to investigate the influence of the spin degree of freedom on the charge currents via the spin-orbit coupling (SOC)[1], a relativistic effect in condensed matter systems. For example, the spin Hall effect causes a transverse spin current induced by a longitudinal charge current[2-4]; whereas the inverse spin Hall effect (ISHE) produces a transverse charge current that is caused by a pure spin current (see Fig. 1a)[5-7]. The ISHE can be used to determine the spin-mixing conductance across ferromagnet/metal interfaces and the spin Hall angle of heavy-atom metals that possess strong SOC[8-14]. In contrast, light-element metals[15], semiconductors[16-18], and organic materials[19,20] are characterized by intrinsically weak SOC, which restricts ISHE detection due to the small spin-current→charge-current conversion efficiency. The technical limitation for these cases stems from the low applicable maximum power for most continuous-wave (cw) microwave (MW) ISHE experiments (<200 mW). Furthermore, small cw ISHE signals are often concealed by potentially confounding spurious signals such as the anomalous Hall effect (AHE), spin backflow, anisotropic magnetoresistance (AMR), etc[21,22]. Thus the formation, manipulation, and detection of spin currents in weak SOC solids has proven to be challenging; but at the same time the focus of recent research referred to in the literature to spin-orbitronics[1].

Pulsed magnetic resonance techniques have been widely employed for the study of spin interactions that affect the optoelectronic properties of organic solids[24,25] using kW-range powers on nanosecond-time scales. Based on previous cw ISHE results and the prediction of current spin pumping models, the ISHE response should increase linearly with the MW power[8,9,21]. During such pulsed excitation of ferromagnetic resonance (FMR)[26,27], a much larger spin current density can be generated that is capable of inducing a large pulse-ISHE (p-ISHE) response that may be measurable even for materials with very weak SOC. In addition, benefiting from the versatility of OSEC synthesis, their SOC can be enhanced by integrating intrachain heavy metal atoms (such as Pt, Ir, etc.)[28,29]. Therefore, by chemical adjusting the heavy metal concentrations in the polymer chains, the SOC strength in OSECs can be systemically tuned[29] and quantitatively investigated by the ISHE response generated from pulsed-FMR (p-FMR).

Here we present the experimental realization of p-FMR for generating and detecting spin currents in OSEC compounds with tunable SOC via p-ISHE. Using this technique we develop methods to effectively eliminate various spurious effects such as AMR, AHE and resonant MW heating that interfere with the p-ISHE current. To highlight the utility of these technical advances, we report a systematic study of p-ISHE response in a variety of OSEC materials with tunable SOC, ranging from π-conjugated polymers (PCP) having strong SOC that contain intrachain Pt atoms with various concentrations, to PCP that are fully organic and thus having very weak SOC, to $C_{60}$ films. We unravel the important role of SOC in spin current→charge current conversion within these materials, and quantitatively determine the value of the spin diffusion length and spin Hall angle in each compound, exploring potential candidate materials for organic spin-orbitronics[30-36].

**I. The pulsed ISHE technique**

Distinct from conventional cw-ISHE devices[8,9], the p-ISHE devices presented here and their electrical connections are deposited as thin film structures[24,25] with slightly altered geometry (see Fig. 1a and Methods). The ferromagnet/OSEC two-layer structure is connected by two small Cu contacts with a gap of 50 μm underneath the device area. The device geometry is chosen such that it places the sample at a position within the cylindrical MW cavity where the $\boldsymbol{B}_1$ field is both maximal and homogeneous, while the $\boldsymbol{E}_1$ field is minimal. The photo in Fig. 1a (inset) displays a finished p-ISHE device and template after it has been installed on the sample rod.

A ferromagnetic (FM) metal of $Ni_{80}Fe_{20}$ (NiFe) is evaporated on top of the spin coated sample film, which is used for FMR that induces spin injection into the adjacent nonmagnetic layers. Cu is chosen as the electrode material since it has negligible SOC, and thus does not interfere with the ISHE from the spin current diffusion through the OSEC film into the underlying electrode[20]. The dynamics of the magnetization vector $\boldsymbol{M}$ under p-FMR induces a pulsed spin current in the adjacent non-magnetic layer by spin pumping. The existence of a pulsed spin current ($\boldsymbol{J}_S$) in materials with a finite spin Hall angle ($\theta_{SH}$) leads to an electric field $\boldsymbol{E}_{ISHE}$, which is referred to as ISHE[5]. $\boldsymbol{E}_{ISHE}$ is transverse to the spin current $\boldsymbol{J}_S$ and spin polarization $\boldsymbol{S}$: $\boldsymbol{J_C} \sim \sigma \boldsymbol{E_{ISHE}} \propto \theta_{SH} \boldsymbol{J_S} \times \boldsymbol{S}$, where $\boldsymbol{J}_C$ and σ are respectively the charge current that originates from the ISHE, and material conductivity along the interface (see Fig. 1a). Therefore, by measuring $\mathbf{V}_{ISHE}$ or $\boldsymbol{J}_C$ in the sample film adjacent to the FM layer, the ISHE can be used for a sensitive detection of the spin Hall angle, which is strongly influenced by the material SOC.

Typical p-ISHE response from a prototype NiFe/Pt system measured at in-plane (i.e. $\boldsymbol{\theta_B}$=0º) external magnetic field, $\boldsymbol{B}$ is shown in Fig. 1b. Pt is used here because of its strong SOC, which generates large ISHE response. The transient voltage was measured between the two Cu contacts underneath the Pt film after one-pulse excitation of 5 μs duration that is averaged over 10240 MW pulses at the resonance magnetic field $\boldsymbol{B}=\boldsymbol{B}_{\text{res}}$ in the FMR spectrum (Fig. 1c). The p-ISHE voltage shows a step function response of ~1.3 mV that occurs simultaneously with the MW pulse. The abrupt $\mathbf{V_{ISHE}}$ rise and decay times are ~5 ns, consistent with the time scale of free induction decay for FMR (see S.I. Fig S1). This indicates that the process of magnon activation, spin current injection, and ISHE generation in Pt metals have all ultrafast response[26,27]. Fig. 1c shows a two-dimensional data set that shows the obtained $V_{ISHE}(B, t)$, whereas Fig. 1d shows $V_{ISHE}(B, t\text{=3 μs})$ response having FWHM $\Delta B$ ~10 mT, which is in agreement with the FMR linewidth from the same device (shown in Fig. 1e).

Comparison between the cw- and p-ISHE responses is performed using the same device at in-plane $\boldsymbol{B}$ (see Methods). The p-ISHE response ($\mathbf{V_{ISHE}}$~1.6 mV) is about two orders of magnitude larger than the cw-ISHE ($\mathbf{V_{ISHE}}$~19 μV); this is also one order of magnitude higher than previously reported ac-ISHE using ac spin pumping[22,23,37,38]. We also confirmed that the p-ISHE response increases linearly with MW power ($\propto B_1^2$) within the MW power used in this work (Fig. 1d inset).

Fig. 1e shows the respective FMR spectra ($dI(B)/dB$) for the NiFe/Pt device and a NiFe-only film before the Pt metal deposition. The FMR linewidth enhancement ($\Delta H_{pp}$) indicates an increased magnon damping, which demonstrates the injection of spin current from NiFe into the Pt layer by spin pumping[39,40]. Figures 1f-1h show the p-ISHE($B$) response and p-ISHE magnitudes measured for the NiFe/Pt device as a function of the out-of-plane angle ($\boldsymbol{\theta_B}$) between the device plane and $\boldsymbol{B}$. With varying $\boldsymbol{\theta_B}$, the p-ISHE signal disappears at $\boldsymbol{\theta_B}$=90º, changing polarity between $\boldsymbol{\theta_B}$=0º and $\boldsymbol{\theta_B}$=180º**,** which is consistent with ISHE($\boldsymbol{\theta_B}$) characteristic response[21]. These results also demonstrate that thermoelectric artifacts[41] are largely suppressed in p-ISHE when using low pulse duty cycle.

The asymmetry of the p-ISHE($B$) response seen in Fig. 1d originates from the extrinsic electromagnetic AHE, which is due to the electric field ($\boldsymbol{E}_1$) component of resonant MW in the NiFe film, when the sample is placed off the resonator center[5,21]. Consequently the obtained ISHE($B$) response can be separated into two components: an actual ISHE($B$) component having a Lorentzian lineshape, and an AHE($B$) component with a dispersive response. Therefore, for the small anticipated ISHE response expected for OSEC materials caused by their weak SOC, the AHE component might eventually dominate the entire response (see for example Fig. 2c). In order to substantially reduce the AHE and other spurious effects in the p-ISHE measurements, especially for OSEC samples, we integrated a MW shunt capacitor layer into the device geometry (see S.I. Figs. S2 and S3). The added capacitive layer ($SiO_2$/Cu) is designed to absorb the electric field component of the MW that crosses the FM layer[42].

Figures 2a-2c compare the $V_{ISHE}(B)$ response with and without the capacitor layer in three different systems: NiFe/Pt/Cu, NiFe/Cu and NiFe/$C_{60}$/Cu, respectively. The symmetric Lorentzian lineshape of the p-ISHE($B$) response in NiFe/Pt with the capacitor layer (bottom panel of Fig. 2a) indicates that the AHE effect is greatly suppressed in this device; the AHE suppression becomes more effective upon increasing the capacitive layer thickness (S.I. Figs. S2 and S3). In Fig. 2b the spectral shape of the NiFe/Cu device with the capacitor shows that Cu indeed is not a good ISHE material, since the dispersive response seen without the capacitor is merely due to AHE($B$) contribution[22]. Figure 2c shows that $V_{ISHE}(B)$ response obtained in an organic layer (50 nm thick $C_{60}$) is relatively weak, so that the AHE component contributes substantially to the response (ISHE/AHE ratio is ~2). Upon suppressing the AHE component with a shunt capacitor (ISHE/AHE ~5), the ISHE component is revealed having $V_{ISHE}$ ~75 μV. We note that if OSEC materials with much weaker SOC would be measured, the AHE component would completely dominate the ISHE response; the shunt capacitor protection is thus essential for obtaining proper $V_{ISHE}(B)$ in an OSEC. We therefore conclude that due to the increased amplitude and absence of spurious effects and artifacts, p-ISHE measurements combined with the capacitor protection is superior for detecting subtle ISHE responses in weak SOC materials.

**II. Pulsed-ISHE studies of various organic semiconductors**

We have measured p-ISHE in a variety of OSEC compounds with tunable SOC values. The generally weak SOC of OSEC materials can be enhanced by incorporating intrachain heavy atoms, such as Pt[28], and it can be tuned by changing the intrachain Pt atoms concentration through incorporation of different organic ligand spacers[29]. Triggered by the enhanced SOC these polymers show substantial phosphorescence (Ph) emission from the lowest triplet state, and consequently their emission spectra contain both fluorescence (FL) and Ph bands[28,29]; this can conveniently serve to provide estimates of the relative SOC strength from the intensity ratio of the Ph/Fl in the electroluminescence (EL) spectra of OLEDs based on these polymers[28].

In Fig. 3a we show the chemical structures of three Pt-containing polymers that we synthesized and study here[29]. It is seen that one spacer unit contains a single phenyl ring, dubbed here as Pt-1; whereas the other spacer unit contains three phenyl rings, dubbed as Pt-3. The Pt-Q polymer has a similar structure as Pt-1, except that the phenyl rings in the chain are replaced by a 5,8-diethynyl-2,3-diphenylquinoxaline unit. The normalized EL spectra of these three polymers (Fig. 3a) show both FL and Ph bands. From the ratio of Ph/FL we conclude that the SOC strength decreases from Pt-1 to Pt-3 to Pt-Q (see Table I)[29]. Aside from Pt-containing polymers with tunable SOC, we also studied several more conventional PCPs with very weak SOC, including poly(3,4-ethylenedioxythiophene):polystyrene sulfonate (PEDOT:PSS), poly(dioctyloxy)phenylenevinylene (DOO-PPV), and Poly[2,5-bis(3-tetradecylthiophen-2-yl)thieno[3,2-b]thiophene] (PBTTT-C14); and also the fullerene $C_{60}$. Their repeat units are shown in the appropriate figure panels of Fig. 4.

In contrast to the inorganic materials used before for spin pumping experiments[15], OSECs generally exhibit rather low charge-carrier mobility. The resulting low conductivity of OSECs yields very high resistance across the two horizontal Cu contacts at a gap of 50 μm (S.I. Fig. S4). For conventional cw-ISHE measurements of OSEC layers[19] $\mathbf{V}_{\mathrm{ISHE}}$ shows a flat response with the material thickness, $d_{\mathrm{N}}$ (S.I., S3 discussion), which makes it difficult to extract a value for the spin diffusion length, which is a critical parameter in deriving the spin Hall angle[21,43,44]. To overcome this difficulty we used charge current detection by essentially shorting the two horizontal Cu contacts with a current preamplifier, which has low internal resistance, $R_{\mathrm{S}}$ (Fig. 3b and S.I. Fig. S4). The generated p-ISHE charge current, $I_{\mathrm{C}}$, passes through the OSEC layer in the vertical direction perpendicular to the NiFe/OSEC interface and along $l$ (see Fig. 3b) and is detected by a current preamplifier. $I_{\mathrm{C}}$ depends on $d_{\mathrm{N}}$ mainly via the thickness dependence of the impedances involved in the measurement (see Methods and S.I., discussion S3), and this dependence provides one way to estimate the OSEC spin diffusion length.

Figure 3c compares the obtained FMR($B$) spectra from the NiFe films as deposited on glass and on a spin-coated Pt-1 polymer film. The broadening of the FMR($B$) response in the NiFe/Pt-1 film indicates the occurrence of spin pumping at the NiFe/OSEC interface (see Methods and S.I. S3 discussion). Figure 3d shows the p-ISHE($B$) response (as a current, $I_S$ between the Cu electrodes) at MW pulse power $P$=1 kW. The maximum $I_S$ occurs at $\boldsymbol{B}_{res}$, and reverses sign when the in-plane magnetic field direction is reversed from $\boldsymbol{\theta_B}$=**0** to $\boldsymbol{\theta_B}$=180º. The $I_S(\boldsymbol{\theta_B})$ sign reversal is consistent with the reversal of the spin current polarization induced by spin pumping[21]. The remaining asymmetry of $I_S(B)$ is due to a residual AHE caused by a finite thickness of the shunt capacitor. Figure 3e shows the expected[19,20,21] linear increase of $I_S$ with the MW power.

Figure 4 shows $I_S(B)$ for the other six NiFe/OSEC devices with various thicknesses at $\boldsymbol{\theta_B}$=0º and $\boldsymbol{\theta_B}$=180º, measured at $P$=1 kW; the ISHE/AHE ratio is also denoted in each panel. When $\boldsymbol{B}_{res}$ is reversed from $\boldsymbol{\theta_B}$=0º to $\boldsymbol{\theta_B}$=180º, all signals reverse polarity, as expected for ISHE. The insets show the respective broadenings of FMR($B$) due to spin-pumping. The accurate calculation of the Gilbert damping factor α is determined from the angular dependence of the FMR linewidth, for which the magnetic field is rotated from in-plane to out-of-plane (S.I. discussion S4 and Table S1)[39,40]. For the NiFe/Pt-3 and NiFe/Pt-Q bilayers, the proper $I_S$ for the ISHE (i.e. after separating the ISHE and AHE components) are ~230 nA and ~150 nA, respectively. When the Pt-containing polymer is replaced by a fully OSEC polymer (i.e. PEDOT:PSS or DOO-PPV), $I_S$ is significantly reduced (69 nA for PEDOT:PSS; 54 nA for DOO-PPV). The ISHE response component was found to scale with the SOC strength of the materials. This explains the weak obtained signals in the fully organic PCPs having very weak SOC. We note our ability to still observe an ISHE response in DOO-PPV, a polymer previously used in spin valve experiments from which spin diffusion length was inferred[45]. This demonstrates that p-ISHE detection provides a highly sensitive method for studying spin pumping into OSECs.

When PBTTT-C14 is used, no p-ISHE current is detected (Fig. 4e), although the broadening of the FMR linewidth indicates that spin pumping indeed occurs at the interface of NiFe/PBTTT-C14 (Table S-I). The absence of a detectable p-ISHE response in PBTTT-C14 is consistent with the previously reported extremely weak SOC[20]. Additionally, this data can serve as a control experiment for the exclusion of measurement artifacts.

Figure 4f shows the p-ISHE response of a NiFe/$C_{60}$ device. Surprisingly, we found that $C_{60}$ film (25 nm thick) exhibits a large ISHE response, in spite of the fact that it only contains carbon atoms with relatively small SOC. We measured an $I_S$ of ~600 nA, which is larger response than in Pt-1. This cannot be explained by a possible artifact, since the obtained ISHE response lacks the usual AHE component; therefore the large ISHE response is a genuine characteristic response of the $C_{60}$ film (see below).

**III. Determination of the spin-Hall angle in OSEC**

The p-ISHE responses observed in this study exhibit similar experimental characteristics as reported for other inorganic and organic materials by spin pumping, such as Pt[5] and PEDOT:PSS[19]. The p-ISHE current dependency on $\boldsymbol{\theta}_{\mathbf{B}}$, the MW power $P$, and the identity of $\boldsymbol{B}_{\mathrm{res}}$ and the FMR($B$) resonance field are all in agreement with the existing understanding of the ISHE phenomenon (Fig. S7). This shows that the pulsed measurements are indeed a consequence of spin current→charge current conversion in the OSEC, generated by spin pumping through the NiFe/OSEC interface. Consequently, the p-ISHE experiments may be used to probe the spin→charge conversion efficiency in various OSEC materials by determining the spin Hall angle, $\theta_{SH}$.

The spin-Hall angle, $\theta_{SH}$ has been measured before in several metals[15], but rarely in OSECs[19,46]. Here we combine the high sensitivity of the p-ISHE detection with measurements of the Gilbert damping in the NiFe/OSEC bilayers to determine $\theta_{SH}$ for various OSEC materials (see Methods, S.I. discussions S3 and S4) as summarized in Table I.

For metallic spin-transport layers with conductivity $\sigma_N$ comparable to that of a ferromagnetic injector $\sigma_F$ the expression for $V_{\mathrm{ISHE}}$ is

$$V(ISHE) = \frac{l\theta_{SHE}\lambda_N \tanh\left(\frac{d_N}{2\lambda_N}\right)}{d_N\sigma_N + d_F\sigma_F}\left(\frac{2e}{\hbar}\right) j_S^0 \qquad (1)$$

where $\lambda_N$ is the spin-diffusion length in the non-magnetic layer; $d_F$ is the FM thickness; $l$ is the FM length parallel to the interface, and $j_S^0$ is the spin current density injected into the non-metallic layer, which can be calculated from the FMR($B$) response (see Fig. 4 insets, and S.I. discussion S4). For the present p-ISHE experiments Eq.(1) has to be modified (see S.I. Eq. (S6)) to account for the various frequency dependent impedances of the experimental set-up (S.I., discussions S3 and S4). Nevertheless it is clear from Eq. (1) (and Eq. S6) that measurements of $V_{ISHE}$ vs. $d_N$ reveal both $\lambda_N$ and $\theta_{SH}$ as long as $d_N \sim < 2\lambda_N$, i.e. when $\tanh\left(\frac{d_N}{2\lambda_N}\right)$ is not saturated at ~1. This is the case for the NiFe/Pt device, where the Pt layer thickness could be controlled within ~1 nm. Eq. (1) (and Eq.(S6)) shows that $V_{ISHE}(d_N)$ first increases with $d_N$; but when $\tanh\left(\frac{d_N}{2\lambda_N}\right)$ saturates, then $V_{ISHE}(d_N)$ decreases approx. as $1/d_N$. This is shown for the NiFe/Pt device in Fig. 5a. We measured $V_{ISHE}(d_N)$ in devices based on several thin Pt layers for which $d_N \sim < 2\lambda_N$, and from the fit to the data using Eq.(1)/Eq.(S6) (Fig. 5a) we obtained independently $\lambda_N(Pt)$ = 2±0.5 nm and $\theta_{SH}$=0.022 (see Table I). These values are in good agreement with the literature values, validating our method and the approximation done in the equivalent circuit (Fig. S4).

For the OSEC layers, especially the Pt-polymers, it is difficult to control their thickness to be within $d_N < 2\lambda_N$~10 nm; in addition, it is possible that at such small $d_N$ the longitudinal transport mode would be tunneling rather than drift. Under these conditions the fabricated OSEC thickness $d_N > 2\lambda_N$ for most of the devices. Consequently $\tanh\left(\frac{d_N}{2\lambda_N}\right) \approx 1$, rendering the two fitting parameters $\lambda_N$ and $\theta_{SH}$ inseparable and thus, according to Eq. (1), their product is treated as one fitting parameter (dubbed here the 'Lambda-Theta product', $\lambda_N\theta_{SH}$). Neither $\lambda_N$ nor $\theta_{SH}$ can be independently obtained from $I_S(d_N)$ when $d_N > 2\lambda_N$. Similar issues for the determination of $\lambda_N$ and $\theta_{SH}$ have been reported for metals such as Pt where values of $\lambda_N$ and $\theta_{SH}$ varied strongly among the different sources, whereas the $\lambda_N\theta_{SH}$ products were well reproduced[11].

Unless $\lambda_N$ can be determined independently from other experiments, the spin Hall angle in OSEC materials in general cannot be determined accurately. One such independent estimate of $\lambda_N$ in OSEC can be made, for example from the giant magneto-resistance (GMR) in organic spin valves (OSV). Although OSV fabrication using OSEC with relatively strong SOC materials is challenging, we nevertheless successfully fabricated OSVs based on $C_{60}$ and Pt-3 polymer in our laboratory, and extracted $\lambda_N$ from the GMR response vs. $d_N$ (Fig. S9 and Table I). We also included in Table I (marked by references) $\lambda_N$ values obtained from literatures, if possible. In addition, we show in Table I values of $\lambda_N$ (marked by *) and $\theta_{SH}$ that we obtained from a method that uses Eq. (S6) keeping $l$ and $j_S^0$ constant, while focusing on the $\tanh\left(\frac{d_N}{2\lambda_N}\right)$ value close to saturation (Fig. S8). For this, we first obtained a crude estimation of $\lambda_N$, and then calculated $\theta_{SH}$. The obtained fits to $J_S$ vs. $d_N$ for Pt-1 and $C_{60}$ devices using the $\lambda_N$ and $\theta_{SH}$ values from Table I are shown in Figs. 5b and 5c, respectively; the good fits support our approach.

The results summarized in Table I show that $\theta_{SH}$ scales with the SOC strength of the OSEC layers; this validates the measuring technique and employed procedure for determining $\theta_{SH}$. We note that an enhancement of the measured $\theta_{SH}$ (compared to the intrinsic $\theta_{SH}^*$ ) occurs in the various OSEC due to their anisotropic conductivity (see S.I. Figs S11 and S12, and S.I. Tables S1 and S2). We found that for all PCPs studied here, except PEDOT-PSS, $\theta_{SH}$ has an opposite polarity compared to that of Pt. Since the spin current→charge current conversion in Pt is mediated by electrons[46], we therefore conclude that holes are the dominant charge carriers in the studied OSECs. PEDOT:PSS is a heavily doped polymer and therefore it is expected that the charge current is carried by free electrons in an impurity continuum band.

Surprisingly, we found that $C_{60}$ films have an anomalously large $\theta_{SH}^*$ which is bigger than that of Pt-1 (but smaller than in Pt). This shows that the p-ISHE experiment is capable of obtaining important information about the SOC even in unusual cases such as $C_{60}$[47]. The π-electrons alone cannot be responsible for the large $\theta_{SH}$ in $C_{60}$ film, since the SOC of these electrons is identical zero. However, carbon σ electrons can also contribute to the SOC in $C_{60}$, because of the mixing that occurs between π and σ electrons due to the strong curvature of the molecule surface[47].

**Summary**:

In summary, we demonstrate a pulsed, high MW power measurement scheme for obtaining ISHE signals 2-3 orders of magnitude stronger compared to previously employed low-power cw experiments. The transient detection also allows for experimental access to the ISHE dynamics with ~5 ns time resolution suggesting that p-ISHE may have potential for fast spin-orbitronics-based logic applications. The p-ISHE device geometry with a shunt capacitor greatly suppresses spurious effects such as AHE, spin backflow, AMR, and MW heating; and enables the use of traditional ferromagnets (e.g. NiFe) instead of the technically more demanding magnetic insulators (such as Yttrium iron garnet)[15,48,49]. Using the pulsed MW resonance technique we demonstrate that the ISHE can be studied in various OSECs with vastly different SOC values; most notably, we obtained a systematic dependence of the ISHE with SOC in a Pt-polymer series, PCPs with very weak SOC, and $C_{60}$ films.

## Methods:

### (i) Device preparation for the p-ISHE measurements

Al thin film electrodes (150 nm) on glass templates (3×50 mm) were fabricated by sputtering and using conventional optical lithography[24,25]. Two Cu contacts with a gap of 50 µm (extended from an Al bottom electrode) were grown by e-beam evaporation through a shadow mask in a glove box integrated vacuum deposition chamber (Angstrom Engineering Inc.), devoted for metal deposition, having a base pressure of $3\times10^{-8}$ Torr. The templates were subsequently transferred into a second glove box that is devoted to OSEC spin coating through an antechamber under nitrogen atmosphere (~0.1 ppm).

The Pt-1, Pt-3, Pt-Q, and DOO-PPV PCPs were synthesized in-house using literature methods[29]. The PEDOT:PSS polymer (Clevios™, P VP AI 4083) was purchased from Heraeous, and the PBTTT-C14 polymer was purchased from Luminescence Technology Corp (Lumtec.) and used without further purification. The polymer/chloroform solutions were spin-coated onto the templates with various spinning speeds (from 1000 to 8000 r.p.m) to obtain different OSEC film thicknesses, followed by a post-annealing procedure (100ºC for 30 mins) in the glove box. The $C_{60}$ powder was purchased from American Dye Source. Inc., and $C_{60}$ films were thermally evaporated onto the template at a rate of 0.5 to 1.0 Å/s._The OSEC coated templates were transferred in a nitrogen atmosphere back to the first glove box with vacuum deposition chamber.

$Ni_{80}Fe_{20}$ ferromagnetic layers (15 nm thickness) were grown by e-beam evaporation through a shadow mask on the spin coated polymer thin films. Without breaking the vacuum, the fabricated structures were transferred with another shadow mask back to the deposition chamber for e-beam evaporation of a $SiO_2$ (150 to 750 nm) dielectric layer and a top Cu thin film (30 nm). All thin film thicknesses were calibrated using a profilometer. The active device area was 0.7mm×1.0mm.

**(ii) p-ISHE measurement set-up**

The p-ISHE measurements were carried out at room temperature in a Bruker ElexSys E580 X-band (~9.7 GHz) pulsed EPR spectrometer equipped with a dielectric resonator (Bruker FlexLine ER 4118 X-MD5). Both cw and pulsed MWs were applied to the p-ISHE templates in presence of a rough vacuum. The purpose of the all-thin film device design has been to ensure that all conducting components are thinner than the MW skin depth at ~9.7 GHz, leaving the device mostly unperturbed by the intense ***E***-fields within the cavity. Fig. 1a illustrates a p-ISHE device on a glass template that was designed specifically to fit in the MW resonator. The position of the template during operation is such that its contact pads are well outside the resonator volume while the actual sample structure at the opposing far end in the center of the resonator. The MW pulse duration time was either 2 μs or 5 μs (chosen depending on the rise-time of the current amplifier being used) at a repetition rate of 500 Hz. The maximum pulsed MW power was ~1 kW resulting in an amplitude $\boldsymbol{B}_1$=1.1 mT at the sample location.

The p-ISHE responses were detected by the induced electromotive force, $\mathbf{V_{ISHE}}$ using a Femto DHPCA 100 for metals, and Stanford Research SRS 570 current-preamplifier for OSEC material (i.e. $I_S$), with bandwidth setting of 100Hz-1MHz. The current amplifier output was connected to the input of a Bruker SpecJet transient recorder (250 MS/s, 8-bit digitizer) that is built into the ElexSys spectrometer. The sensitivity of the current-preamplifier was chosen to be $10^{-3}$ A/V (Femto DHPCA-100) or 20 μA/V (SRS 570). The p-ISHE($B$) response measurements and time dynamics required averaging over 10240 shots. For the cw-ISHE measurement we used cw MW at power of 200 mW applied to the same resonator and $\mathbf{V_{ISHE}}$ (as a derivative spectrum) by magnetic field modulation and lock-in amplification. The $\mathbf{V_{ISHE}}$($B$) spectrum is converted to a voltage amplitude by numerical integration. The parallel capacitance and resistance in the devices were measured using an Agilent E4980A precision LCR meter. The out-of-plane and in-plane conductivities for various materials were measured by a Keithley 4200 at room temperature.

### (iii) Spin Hall angle calculation

The observed p-ISHE responses enable us to calculate the spin hall angles ($\theta_{SH}$) for various OSEC materials. Here we quantify the $\theta_{SH}$ based on a phenomenological model[21] and equivalent circuit model of our set-up (see S.I., Fig. S4). The p-ISHE response is measured during each MW pulse excitation (5 μs duration) at 500 Hz repetition rate. Consequently, the generated p-ISHE response contains a wide bandwidth of AC signals (from ~100 Hz to ~1 MHz, see S.I. discussion S3). The capacitance of each OSEC film, $C_N$, also needs be considered in the circuit model (Fig. S4). By taking into account the device structure, detection electronics, and the AC response of each electronic component, a simplified expression may be written:

$$I_S(pISHE) = Re[(I_C + I_F)\frac{R_F}{R_S^{SUM}+R_F+\frac{2R_N^{SUM}}{1+i(\omega_j C_{N(j)} R_{N(j)})^{SUM}}}], \quad (2)$$

where $R_F$ and $R_S^{SUM}$ are the series resistances of the NiFe film and current-preamplifier impedance, respectively. $\omega_j$ , $C_{N(j)} = \frac{\varepsilon_{N(j)} w(\frac{l}{2})}{d_N}$ and $R_{N(j)} = \frac{d_N}{\sigma_{N(j)} W(\frac{l}{2})}$, are respectively the j- frequency component (established by the finite Fourier transform), parallel capacitance, and resistance of the organic layer at $\omega_j$ (see S.I. discussion S3 for the derivation). The variables $\varepsilon_{N(j)}$ and $\sigma_{N(j)}$ are the dielectric constant and conductivity of the OSEC material at $\omega_j$. $R_N^{SUM}$ and $(\omega_j C_{N(j)} R_{N(j)})^{SUM}$ are the respective sum of parallel resistance, and product $\omega C_N R_N$ terms averaged over the entire frequency range of the measurement apparatus (Fig. S5 and S6). The parameter $w$ is the width of NiFe layer, whereas $l$ is the length of NiFe thin film. The currents $I_C$ and $I_F$ are the generated ISHE responses at the NiFe/OSEC interface, and AHE response from the NiFe thin film, respectively. The latter response is greatly suppressed by the MW shunt capacitor incorporated into our devices (Fig. 2), but not completely eliminated. The spin-pumping related $I_C$ through the OSEC layer can be expressed as[21]:

$$I_C = l\theta_{SH}\left(\frac{2e}{\hbar}\right)\lambda_N \tanh\left(\frac{d_N}{2\lambda_N}\right) j_S^0\ , \tag{3}$$

where $\theta_{SH}$ and $\lambda_N$ are the respective spin Hall angle and spin diffusion length in the OSEC, and $j_S^0$ is the spin current in the OSEC perpendicular to the NiFe/OSEC interface and along $l$[21]. By measuring $I_C$ dependence on the OSEC thickness, $d_N$ at fixed $l$ and $j_S^0$ we can obtain $\lambda_N$ from a normalized version of Eq. (3) (see Fig. 5 and Fig. S8). The spin current $j_S^0$ is obtained from the attenuation of the FMR response (i.e. FMR($\boldsymbol{\theta}_{\mathbf{B}}$) resonant field and spectral width dependencies, see S.I. Figs. S10). The Lamda-theta product ($\lambda\theta_{SH}$) can be accurately calculated by substituting the above parameters into Eqs. (2) and (3)[21].

## Acknowledgments

The work was supported by the National Science Foundation-Material Science & Engineering Center (NSF-MRSEC grant # DMR-1121252). We acknowledge NSF grant #1404634 for supporting the ISHE measurements in materials with tuned SOC.

## Author contributions

D.S., K.J.S, C.B. and Z.V.V. conceived this study and the experiments. D.S. fabricated the devices. K.J.S., D.S., and H.M. implemented the p-ISHE set up. K.J.S., H.M., and M.K. measured the p-ISHE; D.S., M.K., and C.Z. measured the device conductivity and capacitance. C.Z. and D.S. measured the Pt-polymers electroluminescence spectra. D.S. did the circuit modelling for the p-ISHE current. C.B. and Z.V.V. were responsible for the project planning, group managing, and manuscript final writing. All authors discussed the results, worked on data analysis and manuscript preparation.

## Figure and Table Legends

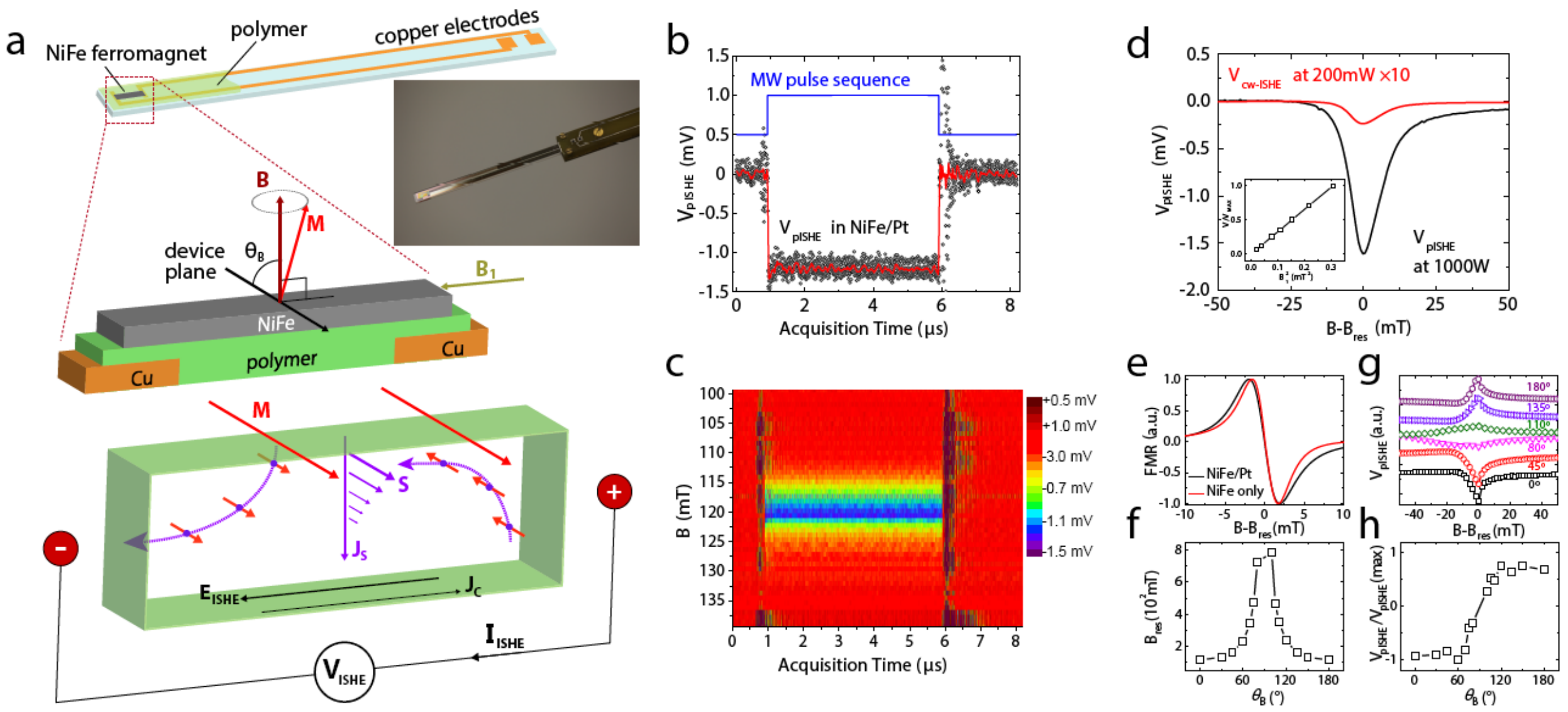

**Figure 1 | Detection of pulsed spin current by p-ISHE. a,** Schematic illustration (not to scale) of the NiFe/OSEC/Cu device on a glass template held by a sample rod with a built-in contact system for thin-film electrical connections[24]. The photo in the right panel shows a polymer-based p-ISHE device fabricated as sketched in the left panel. $\boldsymbol{B_0}$, $\boldsymbol{B_1}$ and $\boldsymbol{M}$ denote, respectively the static external magnetic field, magnetic component of the pulsed MW field, and the dynamic magnetization in the NiFe film that precesses around $\boldsymbol{B_0}$. $\boldsymbol{J_S}$, $\boldsymbol{S}$, $\boldsymbol{J_C}$, $\mathbf{V_{ISHE}}$ and $\mathbf{I_{ISHE}}$ denote, respectively the flow of the pulsed spin current, spin polarization vector, generated electric current, p-ISHE voltage and detected p-ISHE current. **b** and **c**, Time (**t**) and field ($\boldsymbol{B_0}$) responses of the p-ISHE voltage measured for the prototype NiFe (15 nm)/Pt (10 nm)/Cu (30 nm) device under 1kW microwave excitation. The blue solid line in **b** shows the one-pulse MW excitation. The red solid line indicates the moving-average ISHE voltage response. The colour plot shows a resonance at $\boldsymbol{B_0}=\boldsymbol{B_{res}}$=120 mT. The two spurious regions outside the MW pulse originate from MW switching artefacts and non-resonant inductive coupling. **d,** Comparison of $V_{p\text{-}ISHE}$ (at 1kW) and maximum cw-ISHE (at 200 mW) response on the same NiFe/Pt/Cu device. The inset shows the MW power dependence response of $V_{p\text{-}ISHE}$. **e to h,** Field ($\boldsymbol{B}$) and angular ($\boldsymbol{\theta_B}$) dependencies of FMR absorption and p-ISHE response, respectively in NiFe/Pt/Cu device. **e**, Comparison of NiFe FMR response before (red) and after (black) the deposition of Pt. **f**, The resonance field as a function of $\boldsymbol{\theta_B}$, as obtained from FMR in the NiFe film. **g**, Normalized $V_{p\text{-}ISHE}(B, \boldsymbol{\theta_B})$ response, where $\boldsymbol{B_{res}}$ is normalized. **h,** Normalized $V_{p\text{-}ISHE}$ amplitude and polarity vs. $\boldsymbol{\theta_B}$.

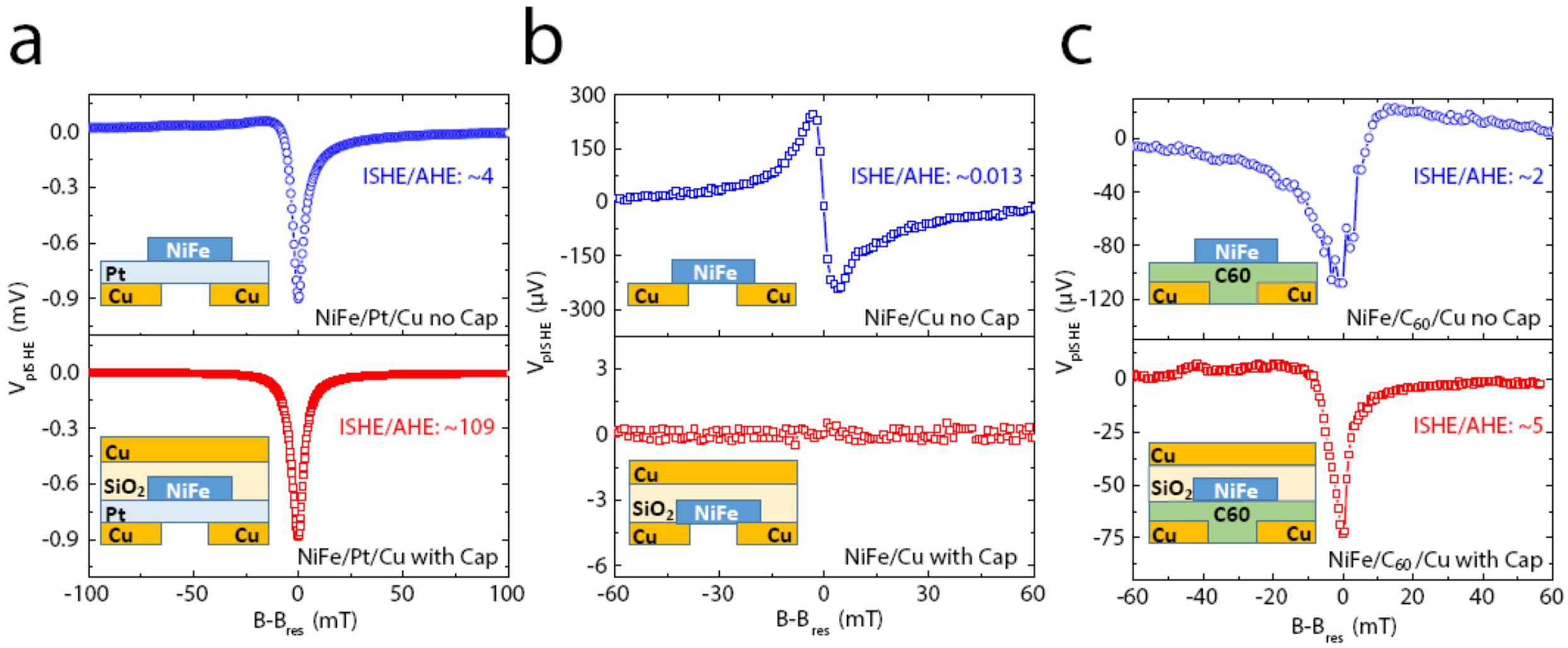

**Figure 2 | Suppression of spurious effects in p-ISHE response using microwave shunt capacitor geometry. a** to **c,** Comparison of p-ISHE responses measured with (red) and without (blue) a $SiO_2$/Cu capacitor layer for (**a**) NiFe/Pt/Cu, (**b**) NiFe/Cu, and (**c**) NiFe/$C_{60}$/Cu devices. The insets are cartoons of the corresponding device geometries with and without the $SiO_2$/Cu capping layers. The respective ISHE/AHE ratios are denoted. The potential spurious effects such as AHE, magnetoresistance, etc. are greatly suppressed in the capacitor geometry. We note, however that the AHE contribution, even with the capacitor protection critically depends on the device alignment in the MW cavity.

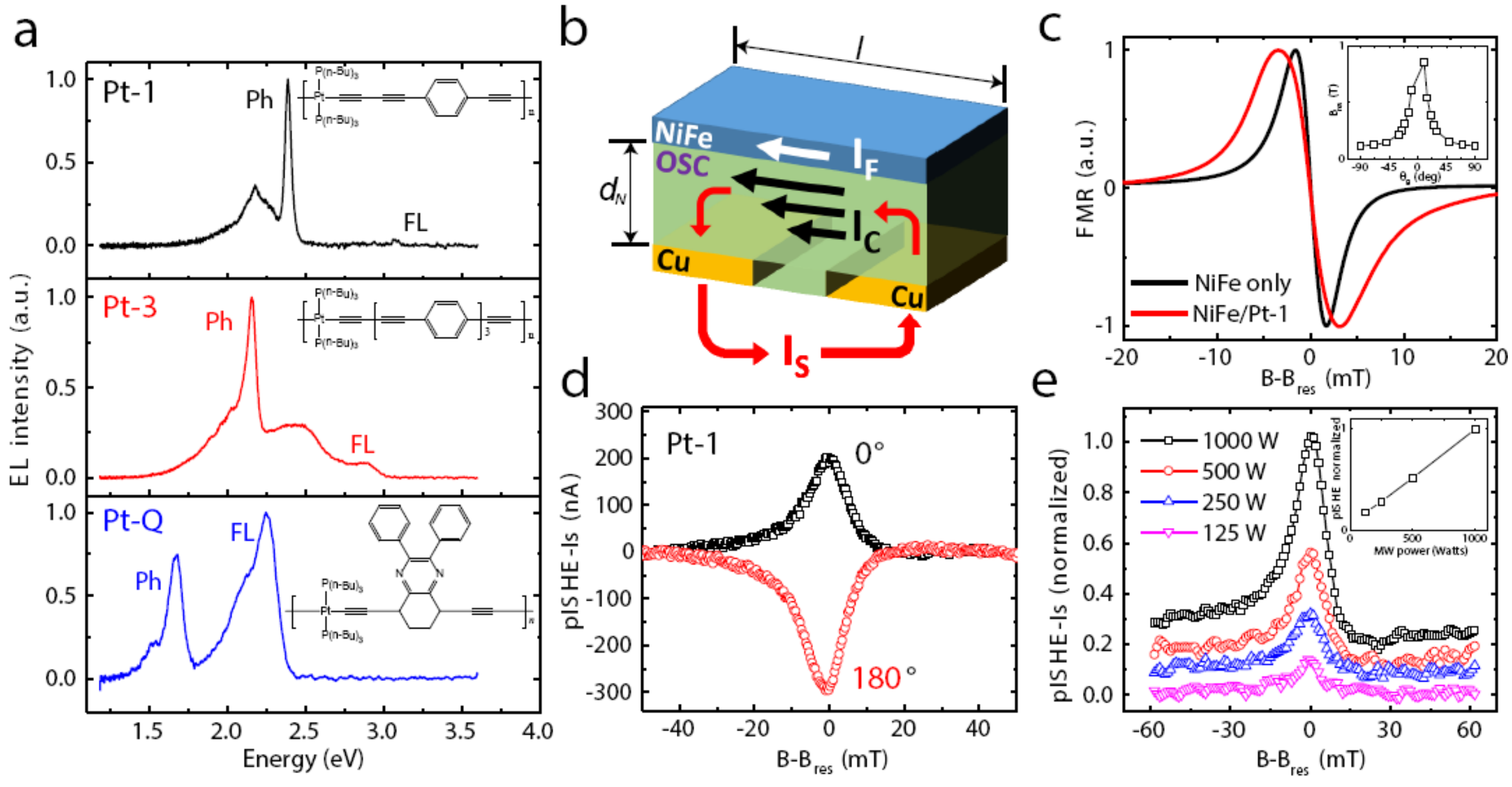

**Figure 3 | The electroluminescence spectra of the Pt-polymer series studied here, and observation of p-ISHE response in Pt-1 polymer. a**, Normalized electroluminescence spectra for Pt-1 (black), Pt-3 (red), and Pt-Q (blue) polymers, respectively. The SOC strengths can be estimated from the electro-phosphorescence (Ph)/fluorescence (FL) intensity ratios[29]. The insets show the building blocks of the studied Pt-polymers. The ‘‘spacer’’ in Pt-1 has a single phenyl ring, whereas that of Pt-3 has three phenyl rings. **b,** schematic illustration for the p-ISHE-*I*s current response in OSEC-based devices. $I_C$, $I_F$, and $I_S$ are respectively the electric current source generated by the ISHE in the organic layer, AHE in the NiFe thin film (suppressed by capacitor geometry), and detected current response by the preamplifier. **c,** FMR spectra of the Cu/Pt-1 polymer/NiFe/$SiO_2$/Cu device measured by MW transmission without (black) and with (red) the spin coated Pt-1 polymer. The inset shows the FMR resonance field, $\boldsymbol{B}_{res}$ vs. the external field angle, $\boldsymbol{\theta}_{B}$. **d,** typical p-ISHE(*B*) response (in terms of current, $I_S$) in Pt-1 polymer device (ISHE/AHE ratio ~9). The black squares and red circles lines in (**d**) are the data with the in-plane magnetic field ***B*** (at 0º) and **–*B*** (at 180º), respectively. **e,** p-ISHE(*B*) response vs. the MW power as denoted. The inset shows the obtained linear $I_S$ vs. MW power dependence.

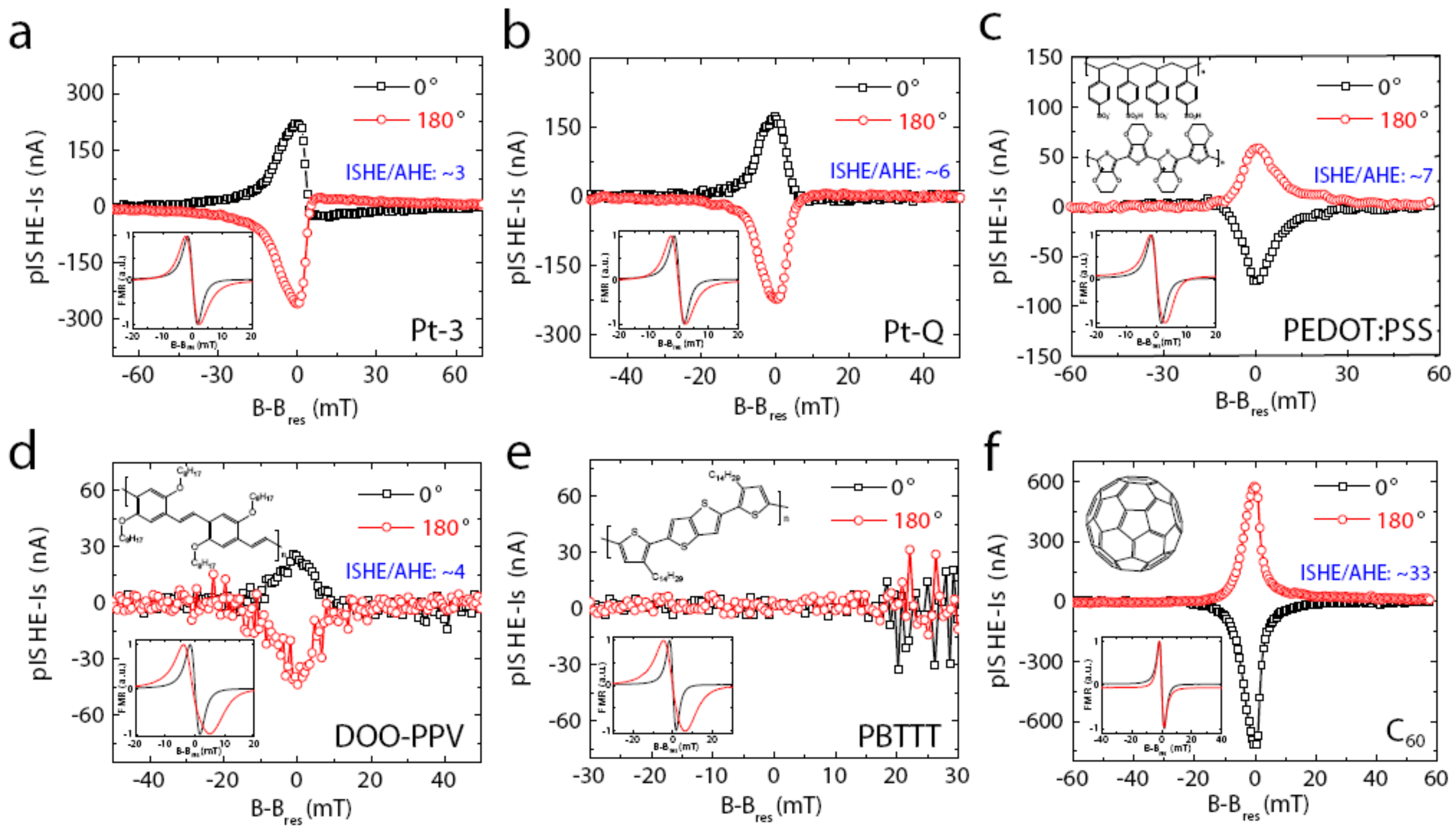

**Figure 4 | p-ISHE(*B*) response in various OSEC materials with tunable spin orbit coupling.** **a** to **f**, p-ISHE(*B*) current response in a variety of NiFe/OSEC/Cu devices as denoted, measured under 1 kW pulsed MW excitation at 10 Hz repetition rate. The OSEC materials are four pristine π-conjugated polymers, PCPs (a, b, d, and e), one heavily doped PCP (c), and a fullerene ($C_{60}$; f). Their respective molecular structures and ISHE/AHE ratios are shown in the appropriate panels. All devices are capped with a $SiO_2$/Cu capacitor layer to suppress the AHE(*B*) response component. The open black squares and red circles are for in-plane field ***B*** (at 0º) and –***B*** (at 180º), respectively. The respective insets show the NiFe FMR(*B*) responses measured by MW absorption in devices with (red) and without (black) the OSEC layer.

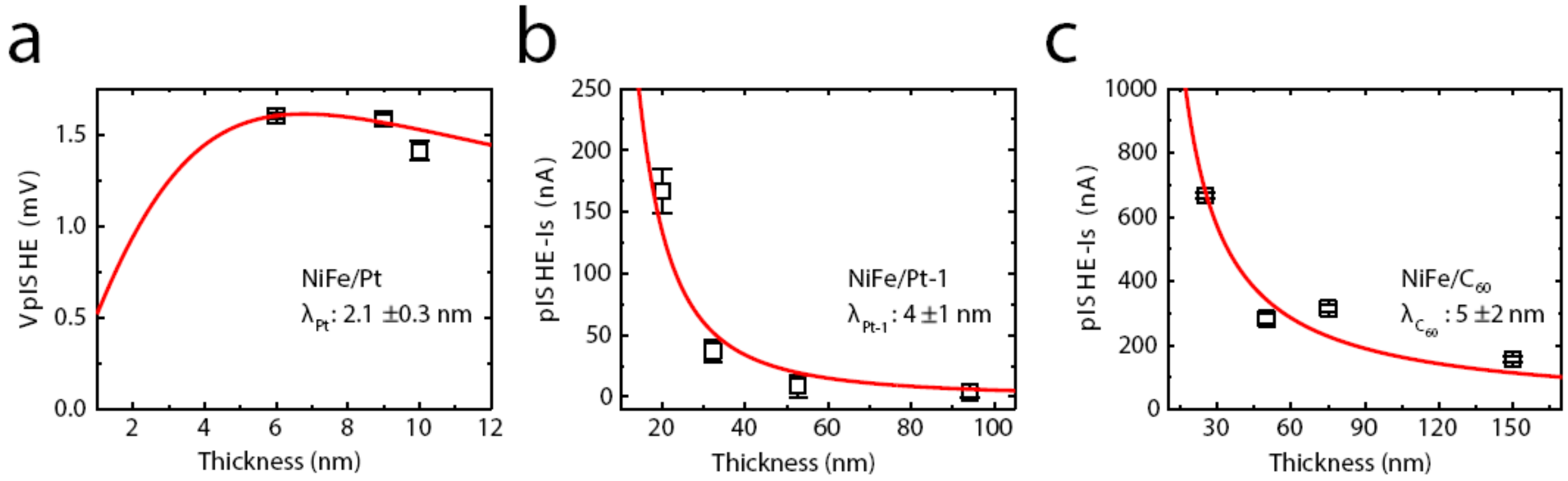

**Figure 5 | p-ISHE(*B*) responses vs. the OSEC thickness. a to c,** thickness dependence of p-ISHE ($I_S$) responses (open squares) in NiFe/Pt/Cu, NiFe/Pt-1/Cu and NiFe/$C_{60}$/Cu, respectively. Red solid lines are fits to the p-ISHE data using Eq.(1) (for Pt) and Eq. (S6) for Pt-1 and $C_{60}$. The respective spin diffusing length extracted from the fit of each OSEC device is denoted.

| Materials | SOC (Ph/FL) | $V_{ISHE}$ (μV) | ISHE-Is (nA) | $\lambda\theta_{SH}$ (nm) | $\lambda$ (nm) | $\theta_{SH}$ |
|---|---|---|---|---|---|---|
| Pt | N/A | 1615 | 2892 | $+7.3\pm0.3\times10^{-2}$ | 2±0.5*, 3.4[11] | $+2.2\pm0.2\times10^{-2}$ |
| Pt-1 | 27 | 76 | 246 | $-1.74\pm0.01\times10^{-3}$ | 4±1* | $-1.2\pm0.3\times10^{-3}$ |
| Pt-3 | 12 | 52 | 231 | $-1.24\pm0.01\times10^{-3}$ | 5±1*, 3§ | $-6.2\pm1.5\times10^{-4}$ |
| Pt-Q | 0.75 | 26 | 145 | $-7.05\pm0.01\times10^{-4}$ | 10±2* | $-7.1\pm1.3\times10^{-5}$ |
| PEDOT:PSS | N/A | 17 | 69 | $+6.6\pm0.6\times10^{-4}$ | 40±10*, 27[19] | $+2.4\pm0.2\times10^{-5}$ |
| DOO-PPV | N/A | 15 | 54 | $-3.29\pm0.01\times10^{-4}$ | 25±10*, 16[45] | $-3.3\pm0.3\times10^{-5}$ |
| $C_{60}$ | N/A | 209 | 668 | $+2.25\pm0.05\times10^{-3}$ | 5±2*, 12[47] | $+4.5\pm1.5\times10^{-4}$ |

*Fitted from thickness dependence of pISHE response by using Eq. (2) and S(6) (see Method and supplementary materials).
§Derived from MR responses in OSVs at low temperature (see supplementary materials).
[Ref]Literatures reported values of spin diffusion lengths.

**Table I: Summary of the important p-ISHE parameters for the investigated OSEC materials obtained from the experiments.** The relative SOC for the three Pt-polymers was obtained from the intensity ratio of the EPH/EL in OLED devices. $V_{p\text{-}ISHE}$ and ISHE-$I_S$ are ISHE voltage and current, respectively between the Cu electrodes measured at $\boldsymbol{B}_{res}$ using MW power of 1 kW. $\lambda\theta_{SH}$ is the 'Lamda-theta product' obtained from the fit to the ISHE response vs. the OSEC thickness using Eqs.(1)-(3) or (S6-S8). λ is the spin diffusion length extracted from Eq. (2) (marked by *), or independent values from GMR response (marked by §), or literatures reports. $\theta_{SH}$ (in radians) is the spin Hall angle calculated from independent spin diffusion length estimations (see S.I. Table S2).

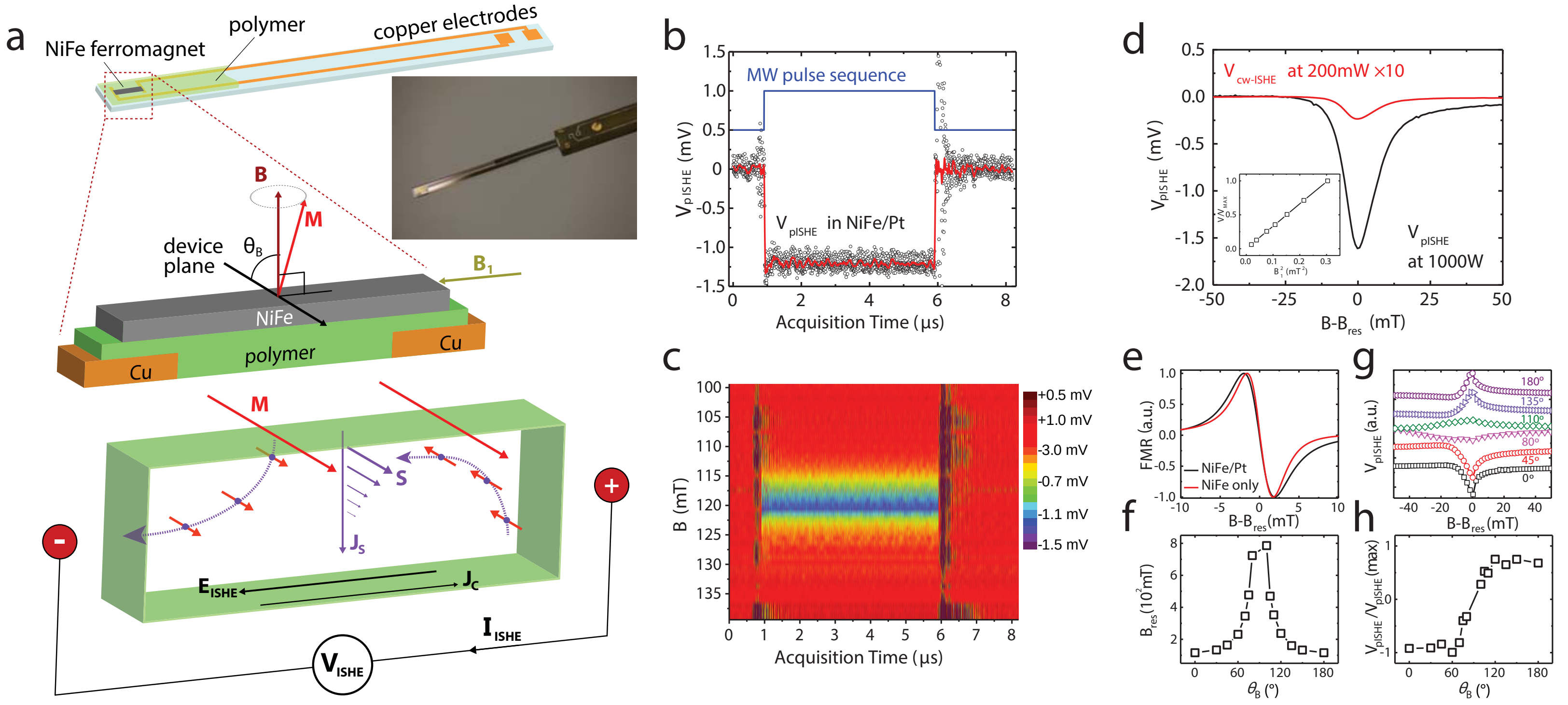

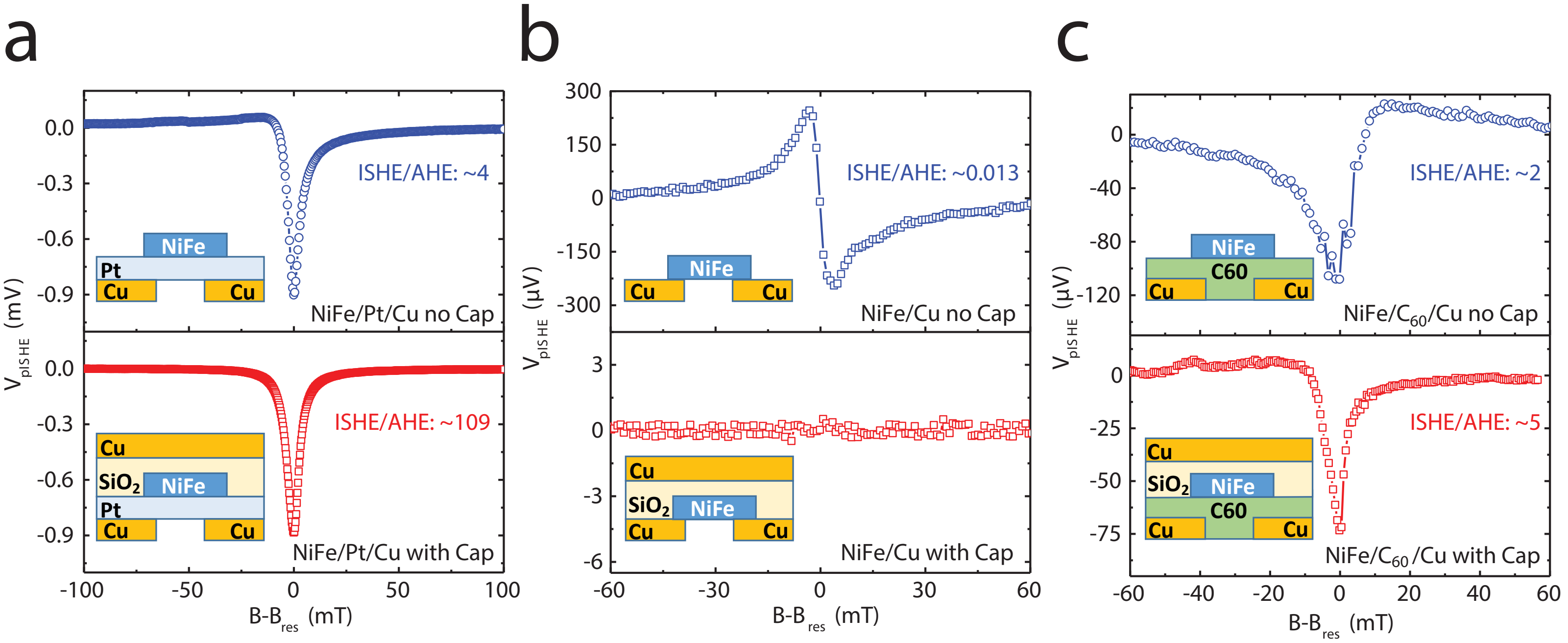

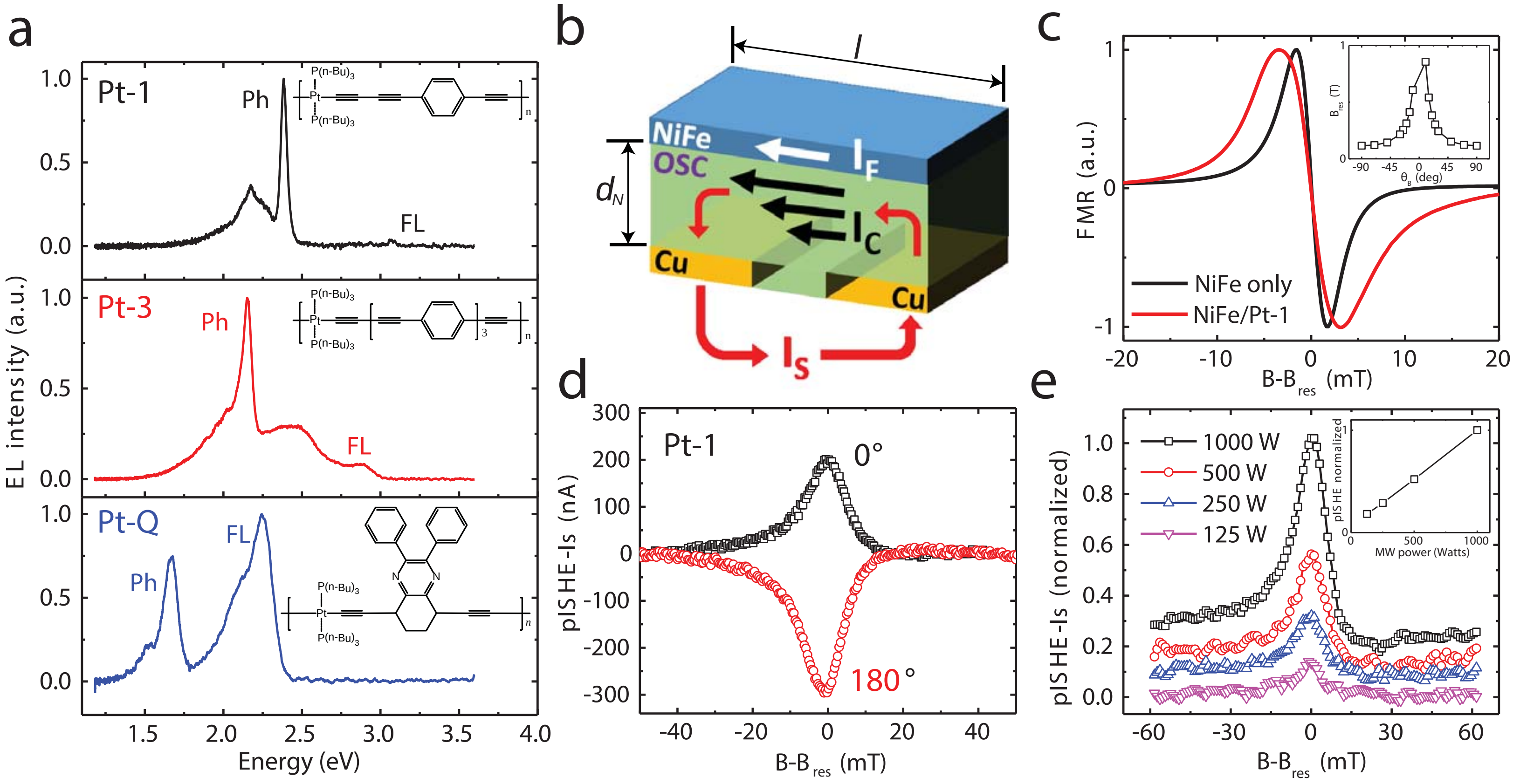

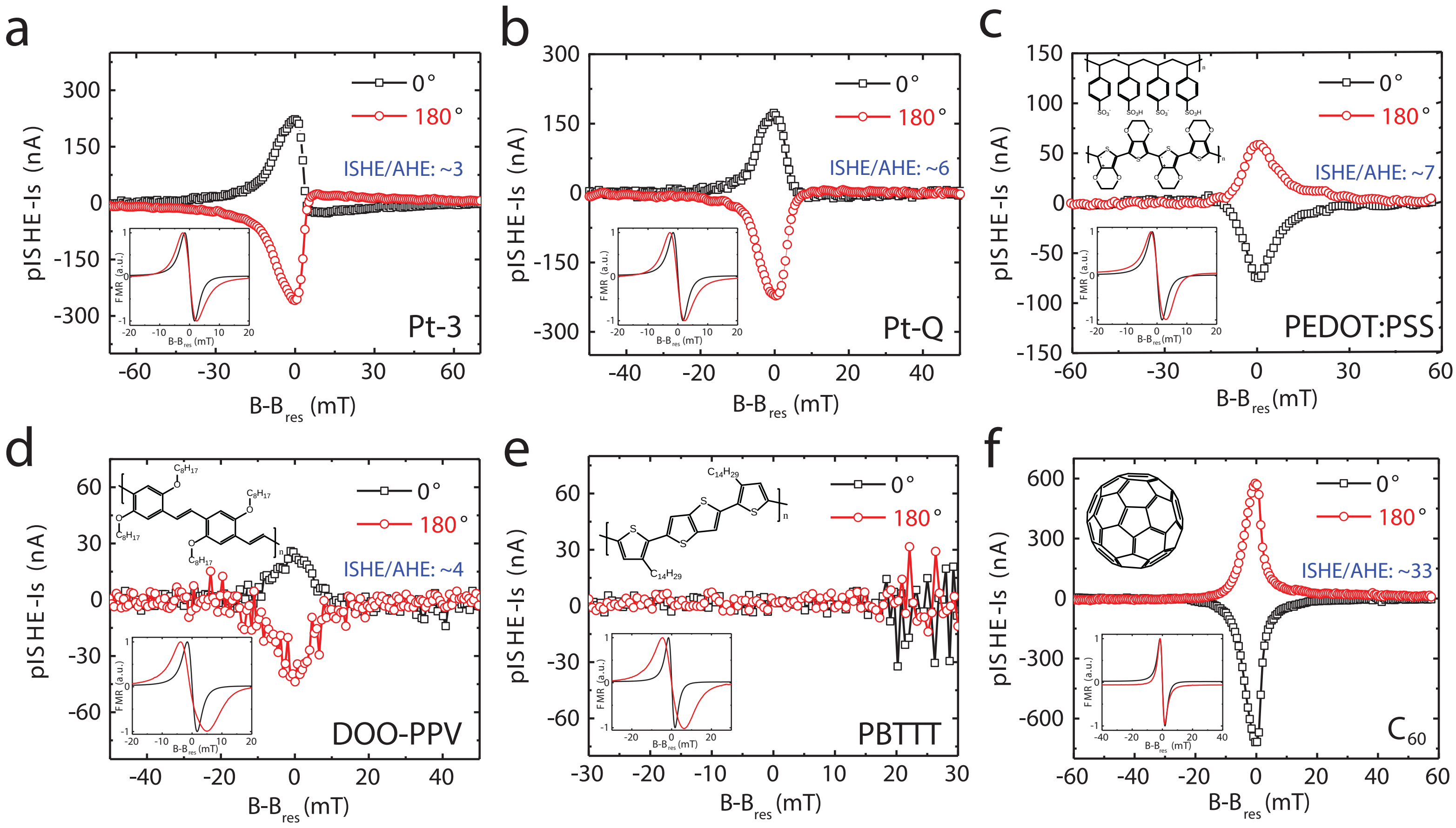

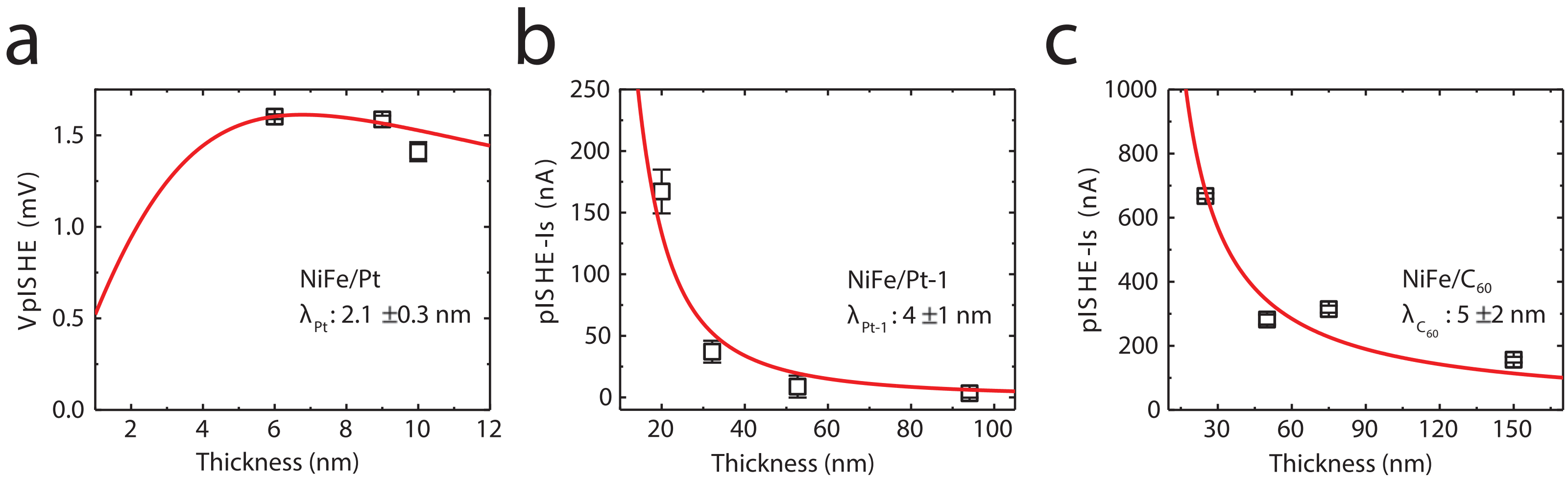

| Materials | SOC (Ph/FL) | $V_{ISHE}$ (μV) | ISHE-Is (nA) | $\lambda\theta_{SH}$ (nm) | $\lambda$ (nm) | $\theta_{SH}$ |
|---|---|---|---|---|---|---|
| Pt | N/A | 1615 | 2892 | $+7.3\pm0.3\times10^{-2}$ | 2±0.5*, 3.4[11] | $+2.2\pm0.2\times10^{-2}$ |
| Pt-1 | 27 | 76 | 246 | $-1.74\pm0.01\times10^{-3}$ | 4±1* | $-1.2\pm0.3\times10^{-3}$ |
| Pt-3 | 12 | 52 | 231 | $-1.24\pm0.01\times10^{-3}$ | 5±1*, 3§ | $-6.2\pm1.5\times10^{-4}$ |
| Pt-Q | 0.75 | 26 | 145 | $-7.05\pm0.01\times10^{-4}$ | 10±2* | $-7.1\pm1.3\times10^{-5}$ |
| PEDOT:PSS | N/A | 17 | 69 | $+6.6\pm0.6\times10^{-4}$ | 40±10*, 27[19] | $+2.4\pm0.2\times10^{-5}$ |
| DOO-PPV | N/A | 15 | 54 | $-3.29\pm0.01\times10^{-4}$ | 25±10*, 16[45] | $-3.3\pm0.3\times10^{-5}$ |
| $C_{60}$ | N/A | 209 | 668 | $+2.25\pm0.05\times10^{-3}$ | 5±2*, 12[47] | $+4.5\pm1.5\times10^{-4}$ |

*Fitted from thickness dependence of pISHE response by using Eq. (2) and S(6) (see Method and supplementary materials).
§Derived from MR responses in OSVs at low temperature (see supplementary materials).
[Ref]Literatures reported values of spin diffusion lengths.

## Supplementary Information

# Inverse Spin Hall Effect from pulsed Spin Current in Organic Semiconductors with Tunable Spin-Orbit Coupling

Dali Sun[§], Kipp J. van Schooten[§], Hans Malissa, Marzieh Kavand, Chuang Zhang, Christoph Boehme*, and Z. Valy Vardeny*

*Department of Physics & Astronomy, University of Utah, Salt Lake City, Utah, 84112, USA*

[§]These authors contributed equally to this work.

To whom correspondence should be addressed: boehme@physics.utah.edu, val@physics.utah.edu

**S1. Time dependence of p-ISHE response in NiFe/Pt device**

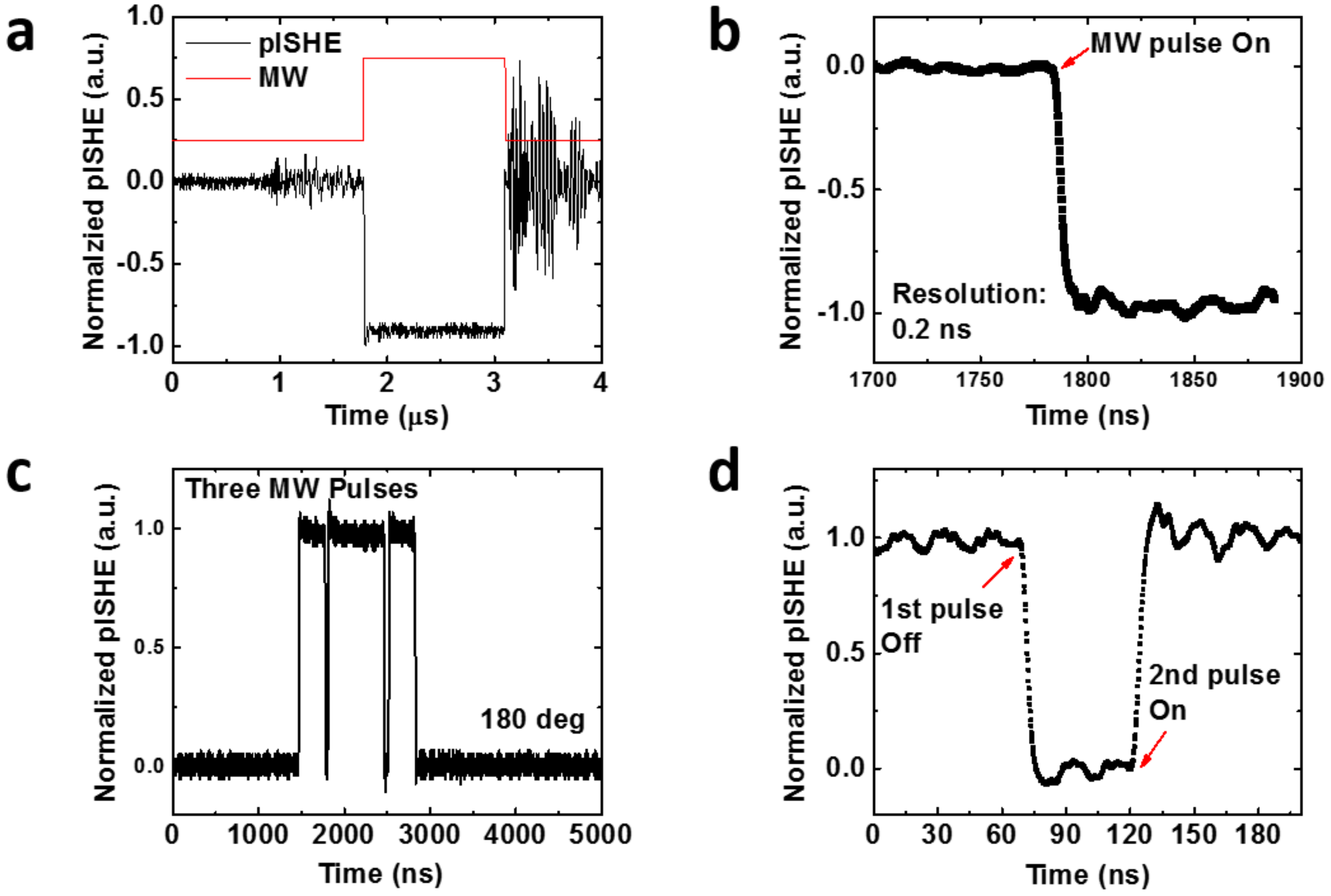

**Supplementary Figure S1 | Time dependence of the p-ISHE response in NiFe/Pt device. a,** normalized p-ISHE response (black solid line) at $\theta_B$ = 0º under a single MW pulse excitation (red solid line) having time resolution of 5 ns. The two spurious regions outside the MW pulse originate from MW switching artefacts and non-magnetic inductive coupling. **b,** Zoom-in on the p-ISHE response with the time resolution of 0.2 ns. **c** and **d,** p-ISHE response at $\theta_B$ = 180º under three MW pulse excitations on a long and short time scale, respectively. The short rise and decay time of the p-ISHE response is consistent with the time scale of the free induction decay in the FMR of the NiFe film. The measurements were obtained at room temperature using a fast oscilloscope.

## S2. Suppressed AHE component as a function of the capacitor layer thickness

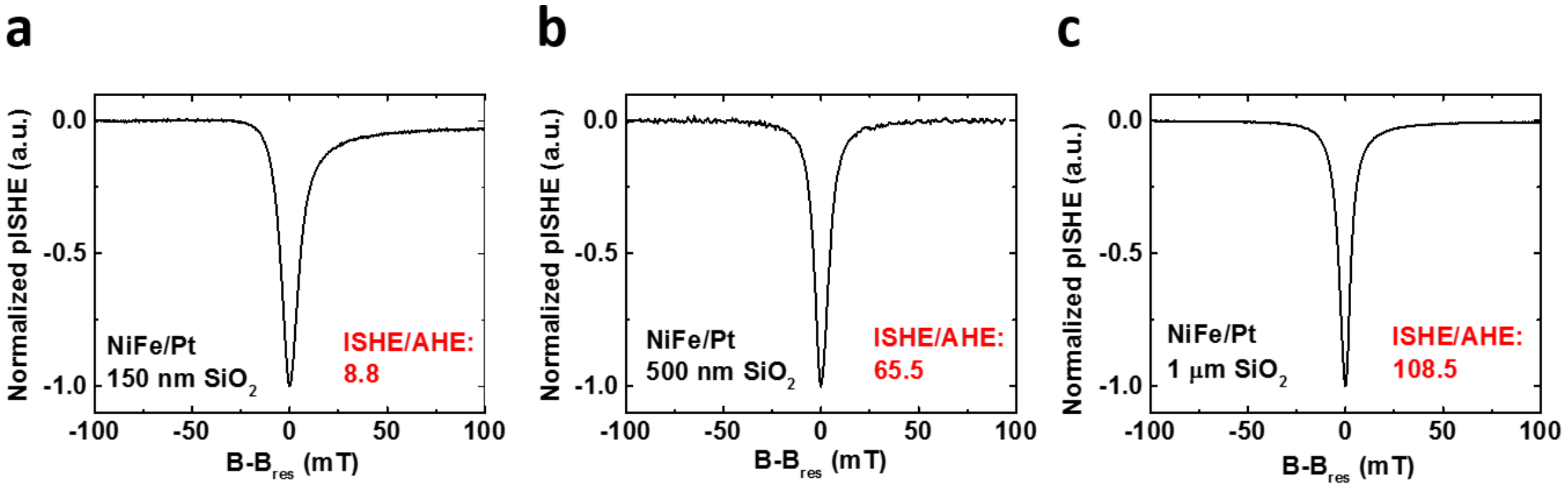

**Supplementary Figure S2 | p-ISHE and AHE components in NiFe/Pt device as a function of the capacitor layer thickness. a** to **c,** p-ISHE($B$) responses in NiFe/Pt devices with $SiO_2$ thickness of 150 nm, 500 nm and 1 µm, respectively. The ISHE/AHE ratio is calculated as shown; note that this ratio increases with the $SiO_2$ layer thickness.

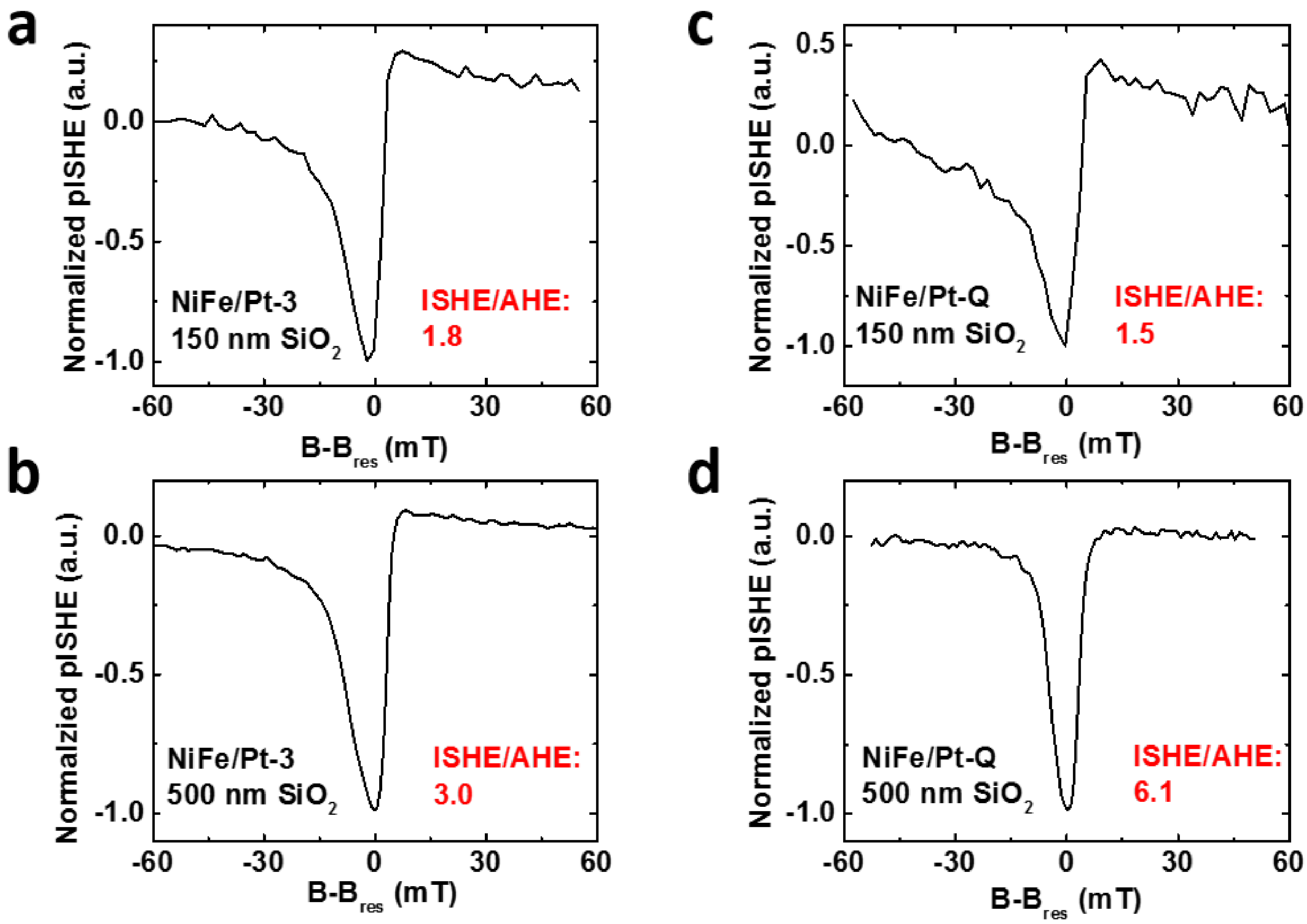

**Supplementary Figure S3 | p-ISHE response in NiFe/Pt-3 and NiFe/Pt-Q devices as a function of the capacitor layer thickness.** **a** and **b,** p-ISHE(*B*) responses in NiFe/Pt-3 capped with $SiO_2$ layer of 150 nm and 500 nm, respectively. **c** and **d**, same as in (a) and (b) but for NiFe/Pt-Q devices. The ISHE/AHE ratio increases with the $SiO_2$ layer thickness. By trading off the depositing time and desired symmetric shape of the p-ISHE response, a 500 nm $SiO_2$ layer thickness was chosen for all NiFe/OSEC devices, whereas a 150 nm $SiO_2$ layer thickness was chosen for the NiFe/Pt devices.

**S3. Description of the p-ISHE response in OSEC materials**

**(i) p-ISHE circuit model**

For the ISHE response in metals under continuous wave (cw) FMR, the induced 'electromotive force' perpendicular to the spin-current direction can be written as[21]

$$V(cwISHE) = \frac{l\theta_{SH}\lambda_N \tanh\left(\frac{d_N}{2\lambda_N}\right)}{d_N\sigma_N + d_F\sigma_F}\left(\frac{2e}{\hbar}\right) j_S^0 \quad \text{(S1)}$$

where $l$ is the length of the NiFe thin film parallel to the NiFe/OSEC interface plane; $\theta_{SH}$ and $\lambda_N$ are the spin Hall angle and spin diffusion length in the OSEC layer, respectively; $d_N$ and $d_F$ are the thicknesses of the OSEC and NiFe layers, respectively; $\sigma_N$ and $\sigma_F$ are the conductivity of the OSEC materials and FM electrode, respectively; $j_S^0$ is the spin current density injected into the OSEC materials at the NiFe/OSEC interface.

In contrast to the DC signals observed in conventional cw-ISHE measurements, the p-ISHE response is a pulse of finite duration that is composed of a range of frequencies. The frequencies' contributions to the induced current $I_S(pISHE)$ are therefore determined by a *discrete* Fourier spectrum (i.e. a Fourier series) that is influenced by the sampling rate and the number of digitized points. For the description of $I_S(pISHE)$ the sample capacitances must be taken into account, in contrast to cw ISHE measurement where these can be discarded. Consequently, equations (S1) used for the cw measurements are no longer applicable for the pulsed ISHE experiments.

For OSEC materials in which $\sigma_F \gg \sigma_N$ and OSEC thicknesses $d_N >> \lambda_N$ (as is the case for the measurements presented here), Eq. (S1) shows that $V_{(ISHE)}$ depends only weakly on $d_N$. The reason is that the denominator is mainly determined by $d_F\sigma_F$ and the $\tanh\left(\frac{d_N}{2\lambda_N}\right)$ term in the numerator approaches unity. The weak dependence on $d_N$ makes it difficult to derive the spin diffusion length in the OSEC layers with this device geometry and experimental setup. Because of this, and for the ferromagnetic NiFe injector, the ISHE current, $I_S$ rather than ISHE voltage is a better choice for detecting the generated ISHE vs. $d_N$. In order to analyze the p-ISHE current we first introduce a circuit model of our set-up as shown in Fig. S4. From the analysis of the circuit model we describe $I_S$ using an equation (S2) that takes into account the impedance created by both capacitance and resistance, when considering the ac-system-response, as follows:

$$I_S(pISHE) = \sum_j^n G_j(\omega_j)\ [(I_C + I_F)\frac{R_F}{R_S+R_F+\frac{2R_{N(j)}}{1+i\omega_j C_{N(j)}R_{N(j)}}}], \quad \text{(S2)}$$

where $R_F$ and $R_S$ are the series resistance in the NiFe thin film and current-preamplifier (taken from the instruments' manual), respectively; $I_C$ is the induced ISHE current close to the OSEC/NiFe interface; and $I_F$ is the current related to the AHE response from the NiFe thin film. The latter component has been greatly suppressed using a MW shunt capacitor in our devices, as shown in Fig. 2 in the text; but not completely eliminated. In Eq. (S2) $C_{N(j)}$ and $R_{N(j)}$ are the measured parallel capacitance and parallel resistance in the OSEC layer at frequency, $\omega_j$; $G_j(\omega_j)$ is the spectral weight of the p-ISHE response at frequency $\omega_j$ (see Fig. S5); and $j$ is the index of the discrete Fourier component.

We consider a simplified expression of Eq. (S2)

$$I_S(pISHE) = Re[(I_C + I_F)\frac{R_F}{R_S+R_F+\frac{2R_N^{SUM}}{1+i(\omega_j * C_{N(j)}R_{N(j)})^{SUM}}}] \quad \text{(S3)}$$

where $\omega_j$, $R_N^{SUM}$ and $(\omega_j C_{N(j)} R_{N(j)})^{SUM}$ are the summation of parallel resistance and $\omega C_N R_N$ term through the entire frequency range available in our set-up (~100Hz to ~1MHz). The corresponding p-ISHE voltage can be then expressed as:

$$V_{pISHE} = Re[I_S\left(R_S + \frac{2R_N^{SUM}}{1+i(\omega_j C_{N(j)}R_{N(j)})^{SUM}}\right)]\ . \quad \text{(S4)}$$

The induced spin current $I_C$ through the OSEC layer can be expressed as[21]

$$I_C = l\theta_{SH}\left(\frac{2e}{\hbar}\right)\lambda_N \tanh\left(\frac{d_N}{2\lambda_N}\right) j_S^0, \quad \text{(S5)}$$

where the parameters were introduced in Eq. (S1). Next, the OSEC layer parallel capacitance and resistance can be estimated: $R_N^{SUM} = \sum_j \left(\frac{d_N}{\sigma_{N(j)} W(\frac{l}{2})}\right)$; $C_N^{SUM} = \sum_j \left(\frac{\varepsilon_{N(j)} W(\frac{l}{2})}{d_N}\right)$, where $W$ is the width of NiFe film; $\sigma_{N(j)}$ and $\varepsilon_{N(j)}$ are the respective conductivity and dielectric constant of the OSEC layer at $\omega_j$. Substituting the expressions for $I_C$, $R_N$ and $C_N$ into Eq. (S2) we get the real part of $I_S$:

$$Re(I_S) = \frac{R_F\left(1+\frac{\omega^2\varepsilon_N^2}{\sigma_N^2}\right)[(R_S+R_F)\left(1+\frac{\omega^2\varepsilon_N^2}{\sigma_N^2}\right)+2\left(\frac{d_N}{\sigma_N*W*(\frac{l}{2})}\right)]}{\left[(R_S+R_F)\left(1+\frac{\omega^2\varepsilon_N^2}{\sigma_N^2}\right)+2\left(\frac{d_N}{\sigma_N*W*(\frac{l}{2})}\right)\right]^2+4\left(\frac{d_N}{\sigma_N*W*(\frac{l}{2})}\right)^2\frac{\omega^2\varepsilon_N^2}{\sigma_N^2}} l\theta_{SH}\left(\frac{2e}{\hbar}\right)\lambda_N \tanh\left(\frac{d_N}{2\lambda_N}\right) j_S^0 \quad \text{(S6)}$$

For analyzing the p-ISHE we need to estimate $j_S^0$. In the model for spin pumping[S1], the injected spin current density $j_S^0$ at the interface is expressed by the relation[21]

$$J_S^0 = \frac{g_r^{\uparrow\downarrow}\gamma^2h^2\hbar\left[4\pi M_S\gamma sin^2\theta_m+\sqrt{(4\pi M_S)^2\gamma^2+4\omega^2}\right]}{8\pi\alpha^2[(4\pi M_S)^2\gamma^2 sin^4\theta_m+4\omega^2]} \quad \text{(S7)}$$

where $\theta_m$ is the magnetization angle to the normal axis of the film plane, $\omega$ is the angular frequency of the magnetization precession (at the MW frequency), $g_r^{\uparrow\downarrow}$ is the mixing conductance, $\gamma$ is the gyromagnetic ratio, $\alpha$ is the Gilbert damping constant, and $M_S$ *is* the saturation magnetization. $h$ is the $\boldsymbol{B}_1$ field component of the MW excitation. The real part of the mixing conductance is given by[21,S2,S3]

$$g_r^{\uparrow\downarrow} = \frac{2\sqrt{3}\pi M_S\gamma d_F}{g\mu_B\omega}(\Delta H_{pp_{(NiFe/OSEC)}} - \Delta H_{pp_{(NiFe)}}); \quad \text{(S8)}$$

Where $g$ is the electron g-factor and $\mu_B$ *is* the Bohr magneton. $\Delta H_{pp_{(NiFe/OSEC)}}$ and $\Delta H_{pp_{(NiFe)}}$ are the FMR spectral peak-to-peak width for NiFe/OSEC) and pure NiFe film, respectively. Now the spin Hall angle $\theta_{SH}$ may be calculated by substituting the above parameters into Eqs. (S2) and (S4)[21].

**(ii) Estimation of the spin diffusing length in the OSEC materials**

Eq. (S6) has been used for fitting the spin diffusion length $\lambda_N$ from the thickness dependence of the p-ISHE response. An alternative is to fit the relative $I_S(d_N)$ dependence to get an estimate of $\lambda_N$. Figure S8 shows the OSEC thickness dependences of p-ISHE-$I_S$ obtained from Pt-Q, DOO-PPV and $C_{60}$. We note that $d_N$ is in most cases much larger than the corresponding spin diffusion lengths, and therefore, the term $\tanh\left(\frac{d_N}{2\lambda_N}\right)$ in Eq. S6 is close to unity. Consequently, the $I_S$ response as a function of $d_N$ is merely dominated by the resistive and capacitive impedance effects in Eq. (S6) that turns out to be ~$1/d_N$. This apparent current decay as a function of the OSEC thickness is thus unrelated to the spin diffusion length. We therefore conclude that $\lambda_N$ estimation using a fitting procedure for $I_S(d_N)$ is accurate only when $d_N$ is sufficiently small such that the term $\tanh\left(\frac{d_N}{2\lambda_N}\right)$ becomes substantially dependent on $d_N$.

For the experiments presented here, thin enough layers ($d_N \sim \lambda_N$ and thus, $\tanh\left(\frac{d_N}{2\lambda_N}\right) < 1$) have been achieved only for Pt and PEDOT. For the other OSECs where $\tanh\left(\frac{d_N}{2\lambda_N}\right) \sim 1$, the spin Hall angle ($\theta_{SH}$) cannot be calculated because $\lambda_N$ and $\theta_{SH}$ always appear as a product, referred here and in the following as Lamda-theta product ($=\lambda_N\theta_{SH}$). We can accurately determine $\lambda_N\theta_{SH}$ from the p-ISHE experiments in these cases, as shown in Table I, yet not the individual parameters $\lambda_N$and $\theta_{SH}$. Nevertheless, we can determine the spin Hall angle for cases where the spin diffusion lengths $\lambda_N$ is known from other experiments such as spin-valve measurements, as is the case for some of the OSEC materials studied here.

We have fabricated organic spin valves (OSVs) based on several OSEC ($La_{0.67}Sr_{0.33}MnO_3$/OSEC/Cobalt/Al); and measured the obtained giant magnetoresistance (GMR) vs. the organic interlayer thickness[30-36] in order to estimate the spin diffusion length independently of the ISHE studies. Fabrication of OSV devices based on DOO-PPV and PBTTT is straightforward since the spin diffusion length in these materials is sufficiently large. In contrast, the SOC in Pt-polymers is much stronger, and therefore it is expected that the spin diffusion length in these polymers is very short. Fabrication of proper OSV devices in these cases has therefore been a challenge. If the Pt-polymer thickness is too small (<5 nm), tunneling magnetoresistance may dominate the OSV response, where the thickness dependence is not related to $\lambda_N$. Also, if $d_N \gg \lambda_N$ then the GMR response may be too small for extracting a reliable value for $\lambda_N$. Nevertheless, we succeeded in fabricating OSVs based on the Pt-3 polymer of which spin-valve results are shown in Fig. S9. The MR response in Pt-3 polymer as function of thickness is given by[30]

$$MR \propto \frac{2P_1P_2exp^{(-d_N/\lambda_N)}}{1-P_1P_2exp^{(-d_N/\lambda_N)}} \quad \text{(S9)}$$

where $P_1$ and $P_2$ are the nominal values for the spin injection polarization degrees of the two ferromagnet (FM) electrodes (namely, P($La_{0.67}Sr_{0.33}MnO_3$, LSMO) ≈95% and P(Cobalt)≈30%)[S4,S5]. Using Eq. (S9) to fit the GMR($\lambda_N$) results we could estimate $\lambda_N$ (see Table I). We note that most OSV responses are only observed at low temperature because of the materials choice for the bottom FM electrode (namely the LSMO)[30]. We thus estimate, as was verified by muon spin rotation measurements that $\lambda_N$ may decrease by a factor of ~2-3[S6] at room temperature. In addition whenever available, spin diffusing lengths obtained from additional independent measurements found in the literature have also been included in Table I. These values permit us to estimate the spin Hall angle from the p-ISHE measurements with uncertainties that are predominantly governed by the uncertainties of the available spin-diffusion lengths.

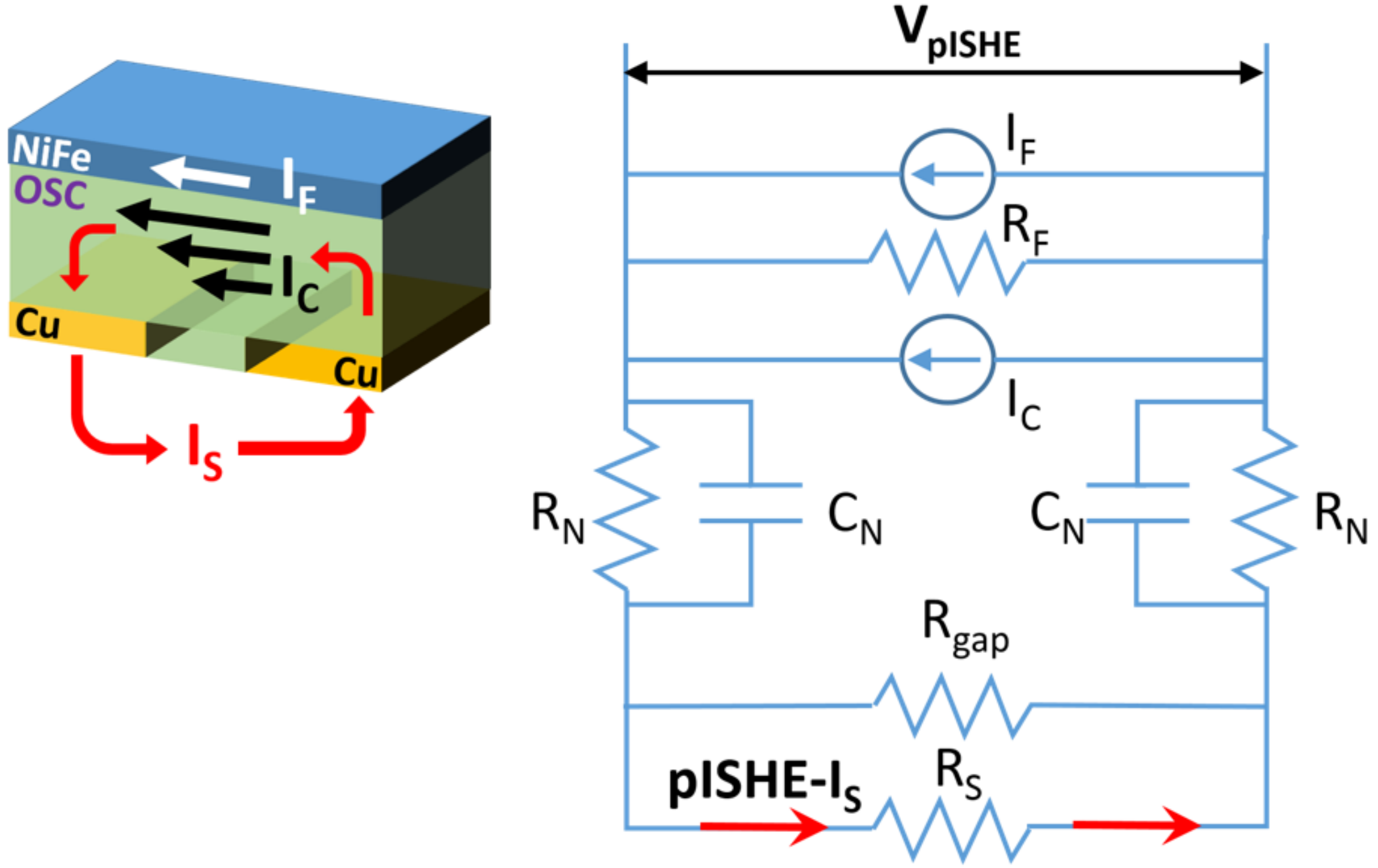

**Supplementary Figure S4 | Schematic illustration and equivalent circuit model for the p-ISHE current response, $I_S$ in OSEC-based devices.** $R_F$, $R_N$ and $C_N$ are the resistance between the OSEC contacts and the NiFe layer, the parallel resistance of the NiFe layer, and the parallel capacitance in the OSEC layer, respectively. $R_S$ is the internal series resistance of the current preamplifier. $R_{gap}$ is the resistance of the organic film between the two Cu electrodes. The conductivities of most OSEC films are low, so that $R_{gap}$ can be considered very large. $I_C$, $I_F$, and $I_S$ are the electric currents due to the ISHE in the OSEC layer, the AHE in the NiFe thin film, and the detected current response by the preamplifier, respectively.

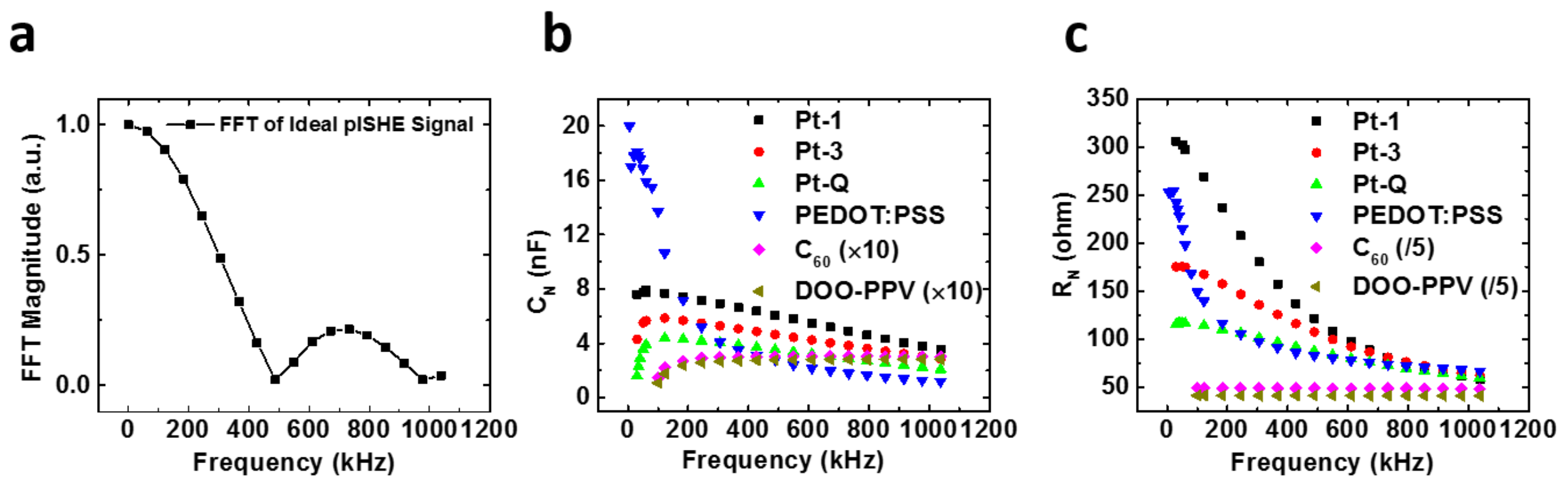

**Supplementary Figure S5 | Frequency dependencies of the p-ISHE, as well as $C_N$ and $R_N$ for various OSEC materials. a,** generated p-ISHE frequency response obtained by Fast Fourier Transform (FFT) from the p-ISHE transient responses. **b** and **c,** measured parallel capacitance, $C_N$ and resistance $R_N$, respectively in various OSEC-based p-ISHE devices.

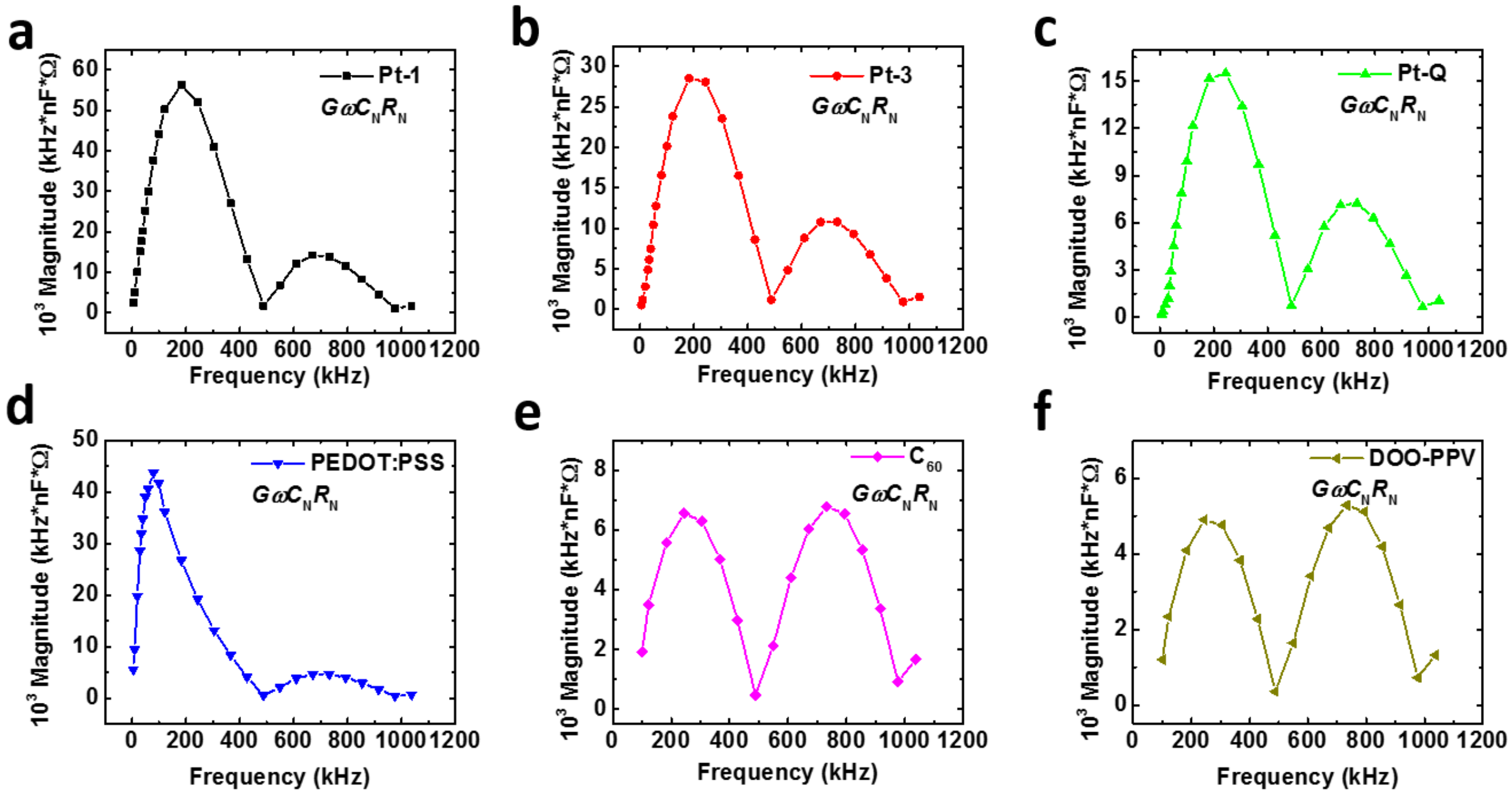

**Supplementary Figure S6 | The frequency dispersion of $C_N$ and $R_N$ in various OSEC p-ISHE devices. a** to **f,** The frequency dispersion of the relation $\omega C_N R_N$ [in Eq. (S3)] in various OSEC-based p-ISHE devices. G is p-ISHE intensity response averaged over the set-up frequency range obtained from Fig. S5(a). The actual values ($\omega C_N R_N$) used in Eq. (S3) for each OSEC material is the summation over the entire frequency range ( $\sum_{j=100Hz}^{j=1MHz} G_j \omega_j C_{N(j)} R_{N(j)}$).

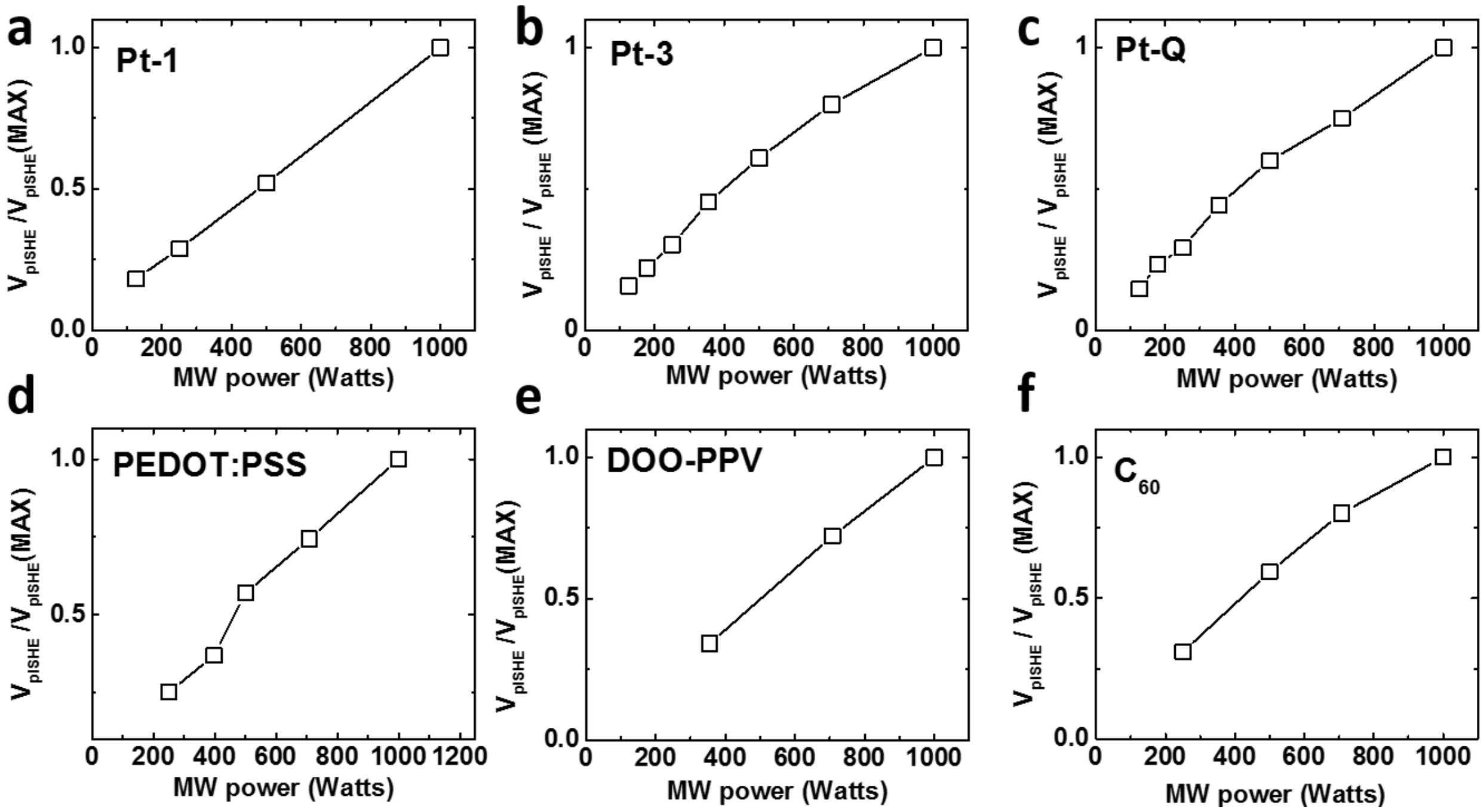

**Supplementary Figure S7 | Dependence of the p-ISHE on the MW power for various OSEC materials. a** to **f,** *I*s due to p-ISHE as a function of the MW power (open squares) for (a) Pt-1, (b) Pt-3, (c) Pt-Q, (d) PEDOT:PSS, (e) DOO-PPV, and (f) $C_{60}$. The connecting lines are a guide for the eye.

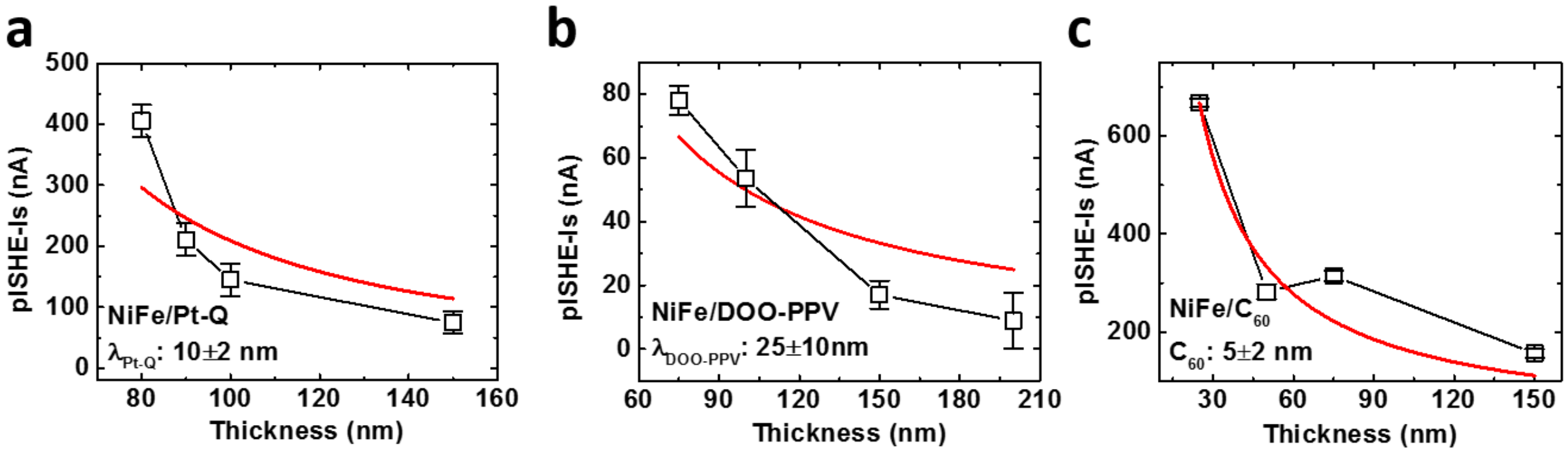

**Supplementary Figure S8 | p-ISHE dependence on the OSEC thickness for various devices.** **a** to **c,** *I*s due to the p-ISHE as a function of the OSEC thickness $d_N$ (open squares) for (a) Pt-Q, (b) DOO-PPV, and (c) $C_{60}$. The connecting lines are a guide for the eye. The red lines through the data points are fits using Eq. (S6). The spin obtained spin diffusion length for each OSEC is given for each OSEC.

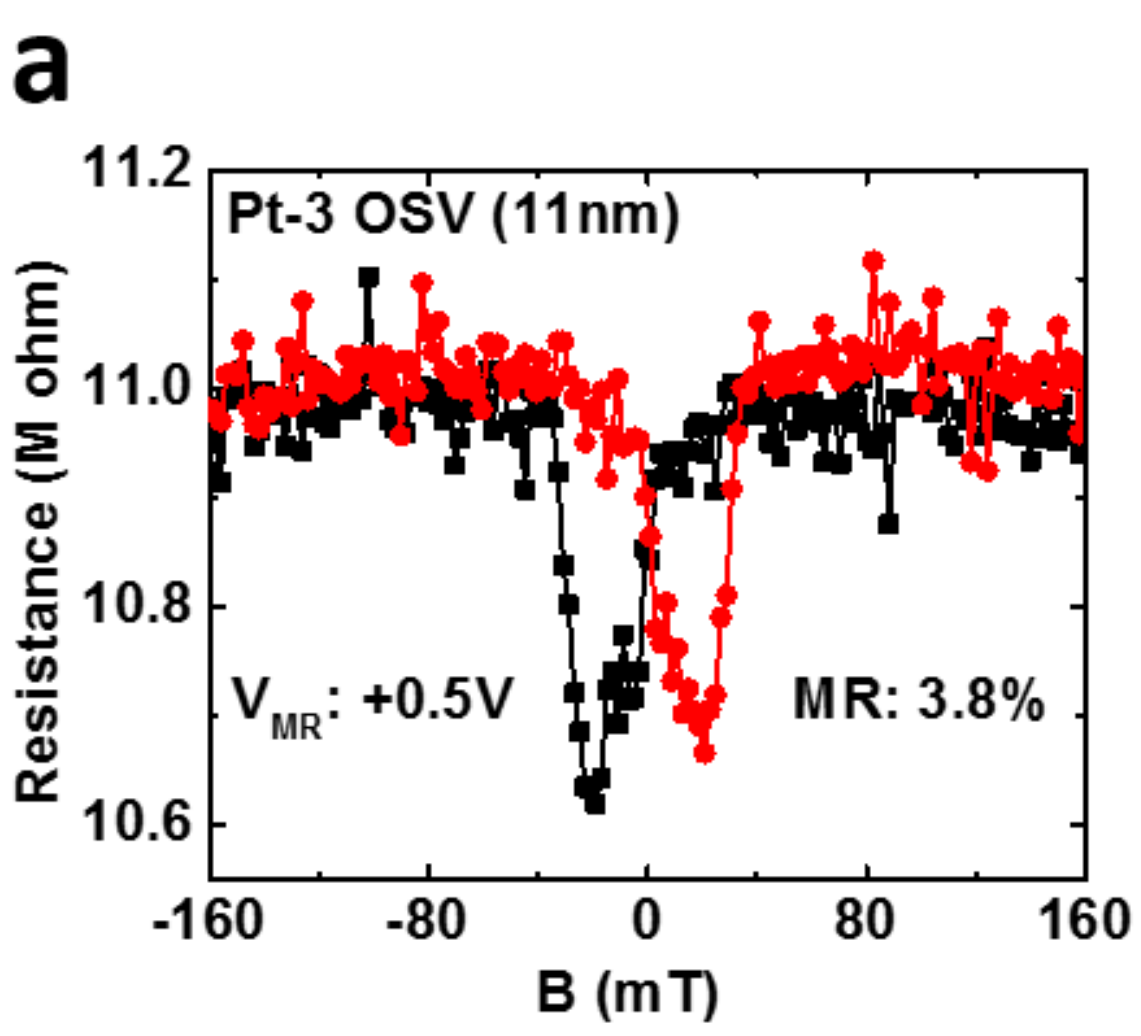

**Supplementary Figure S9 | GMR($B$) response for various Pt-3 OSV devices. a** and **b,** **G**MR($B$) response in Pt-3 OSV device having $d_N$ =11nm and $d_N$ =18nm, respectively, measured at 10K. The spin diffusion length of the Pt-3 polymer at 10K obtained from these measurements is estimated to be ~3 nm.

## S4. Angular dependence of FMR response

In order to calculate the spin Hall angle for various OSEC materials using Eqs. (S5) to (S9), it is necessary to determine $\theta_m$, $\omega$, $g_r^{\uparrow\downarrow}$, $\gamma$, $\alpha$, and $M_S$ in order to estimate the spin current density $j_S^0$. For this purpose the out-of-plane angular dependence of the FMR response was measured for various OSEC materials and subsequently a numerical analysis of these data based on the Landau-Lifshitz-Gilbert (LLG) equation (S10) was made;

$$\frac{dM(t)}{dt} = -\gamma M(t) \times H_{eff} + \frac{\alpha}{M_S} M(t) \times \frac{dM(t)}{dt} \qquad \text{(S10)}$$

Under static equilibrium and neglecting the magneto-crystal anisotropy, the LLG equation yields the expression:

$$2H sin(\theta_H - \theta_m) + 4\pi M_S sin2\theta_m = 0, \qquad \text{(S11)}$$

that relates the external magnetic field angle $\theta_H$ and magnetization angle $\theta_m$ with respect to the normal axis of the NiFe film plane. Here $\theta_H = \theta_B - 90$. $H$ is the strength of the external magnetic field. The FMR resonance condition is given by[S7,S8]:

$$\omega/\gamma = \sqrt{H_1 H_2}; \qquad \text{(S12)}$$

$$H_1 = H_{res}\cos(\theta_H - \theta_m) - 4\pi M_S cos2\theta_m; \qquad \text{(S13)}$$

$$H_2 = H_{res}\cos(\theta_H - \theta_m) - 4\pi M_S cos^2\theta_m; \qquad \text{(S14)}$$

where $H_{res}$ is the obtained FMR resonance field. By numerically fitting the FMR resonance field as a function of the out-of-plane angle (Fig. S10) using Eq. S10 to S12, we can determine the values of $\theta_m$, $\omega/\gamma$, and $4\pi M_S$.

The full-width at half-maximum (FWHM) of the FMR($H$) response (which can be obtained by the peak to peak field difference, $\Delta H_{pp}$ in the derivative FMR($H$) response) is intrinsically caused by the Gilbert damping constant ($\alpha$) (Eq. (S10). It can be expressed as[S7,S8]

$$\Delta H_{in} = \alpha(H_1 + H_2)\left|\frac{d(\omega/\gamma)}{dH_{res}}\right|^{-1}. \qquad \text{(S15)}$$

The FWHM may also be obtained from the anisotropy dispersion in the out-of-plane direction[S9-S12], which is given by

$$\Delta H_{ex} = \left|\frac{dH_{res}}{d(4\pi M_S)}\right| \Delta(4\pi M_S) + \left|\frac{dH_{res}}{d(\theta_H)}\right| \Delta\theta_H. \tag{S16}$$

The extrinsic $\Delta H_{ex}$ originates from the local variation of the magnitude and direction of $4\pi M_S$ at the resonant field ($H_{res}$). Thus $\Delta H_{pp}$ can be expressed as[S9-S12].

$$\Delta H_{pp} = \frac{\Delta H_{in}}{\sqrt{3}} + \frac{\Delta H_{ex}}{\sqrt{3}}; \tag{S17}$$

$\Delta H_{pp}$ vs. $\theta_H$ (right panels of Fig. S10) is fitted using Eq. (S15) to (S17) by adjusting the value of $\alpha$, $\Delta(4\pi M_S)$ and $\Delta\theta_H$.

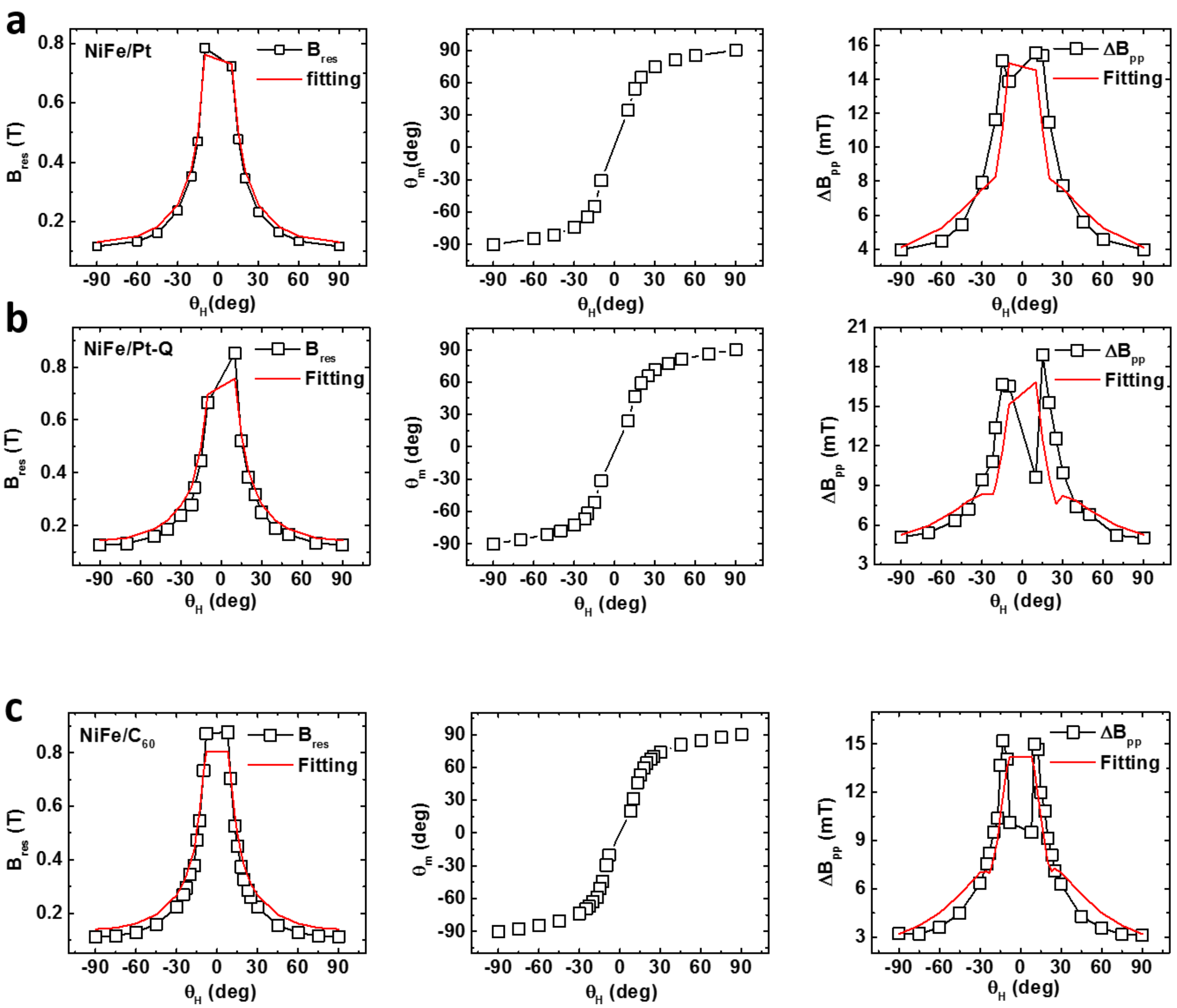

**Supplementary Figure S10 | a** to **c,** FMR resonance field (left panels), magnetization angular dependence $\theta_m$ (middle panels), and FMR width as a function of $\theta_H$ (right panels) in various OSEC-based p-ISHE devices as given. The resonance field as a function of $\theta_H$ is fitted using Eq. (S10) to (S14) (red line) to obtain the parameters $4\pi M_S$ , $\omega/\gamma$ and $\theta_m$. The peak-to-peak width as a function of $\theta_H$ is fitted using Eq. (S15) to (S17) to obtain the damping factor $\alpha$, variation in the value of $4\pi M_S$ (i.e. $\Delta 4\pi M_S$) and angle ($\Delta\theta_H$) for the magnetization at the resonant field. The derivation of the fit around $\theta_H = 0$ is due to the small number of data points at higher resonance fields, a limitation caused by the maximum of the external magnetic field in the pulsed EPR spectrometer ($B_{res}$ has to be < 1T).

## S5. Conductivity anisotropy between out-of-plane and in-plane direction in the studied OSEC materials

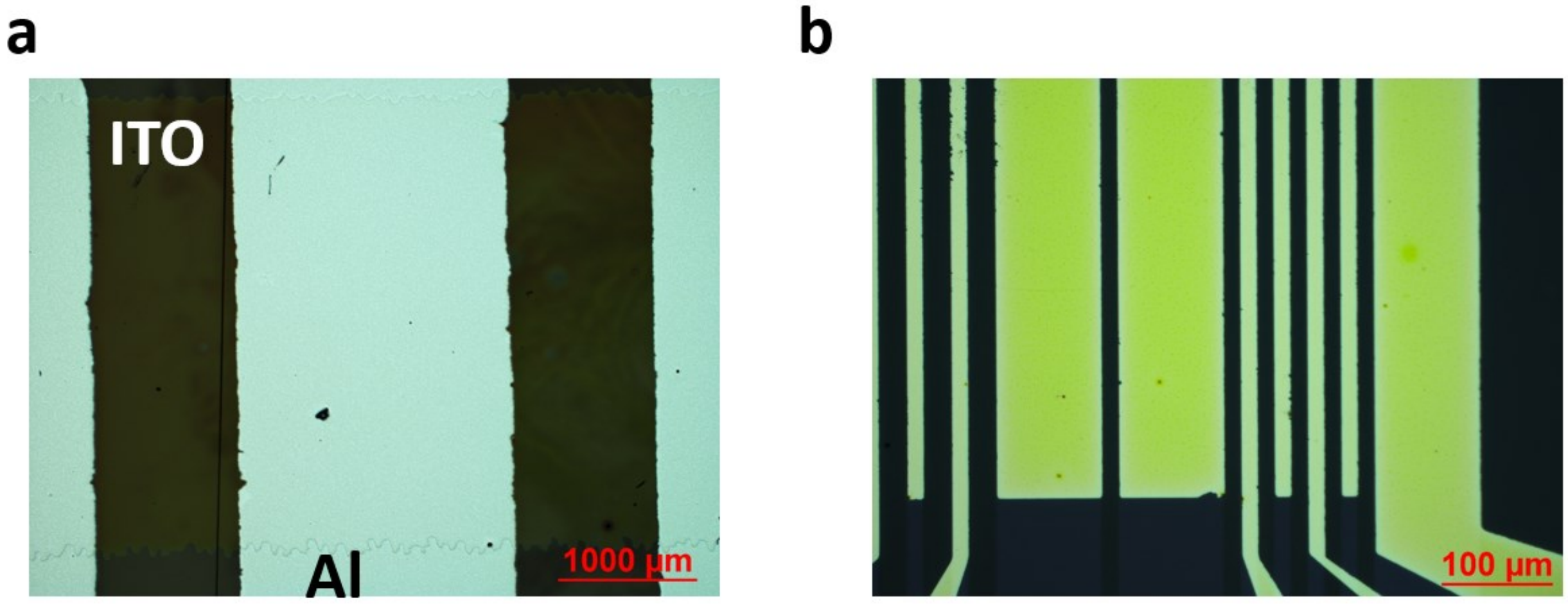

**Supplementary Figure S11 | Device geometry for conductivity measurements of the OSEC materials used in this study. a,** The OSEC out-of-plane conductivities were measured using an organic light-emitting diode (OLED) structure. The vertical dark stripe is part of the bottom-ITO anode on which the OLED is fabricated. The vertical light-blue stripe is the top Ca/Al cathode. The OSEC layer sits in between the two electrodes. The device area is 2.5 mm × 2.0 mm. **b,** In-plane conductivities were measured by a four-probe stage (Keithley 4200). The gap between the electrodes (dark) is 15 μm.

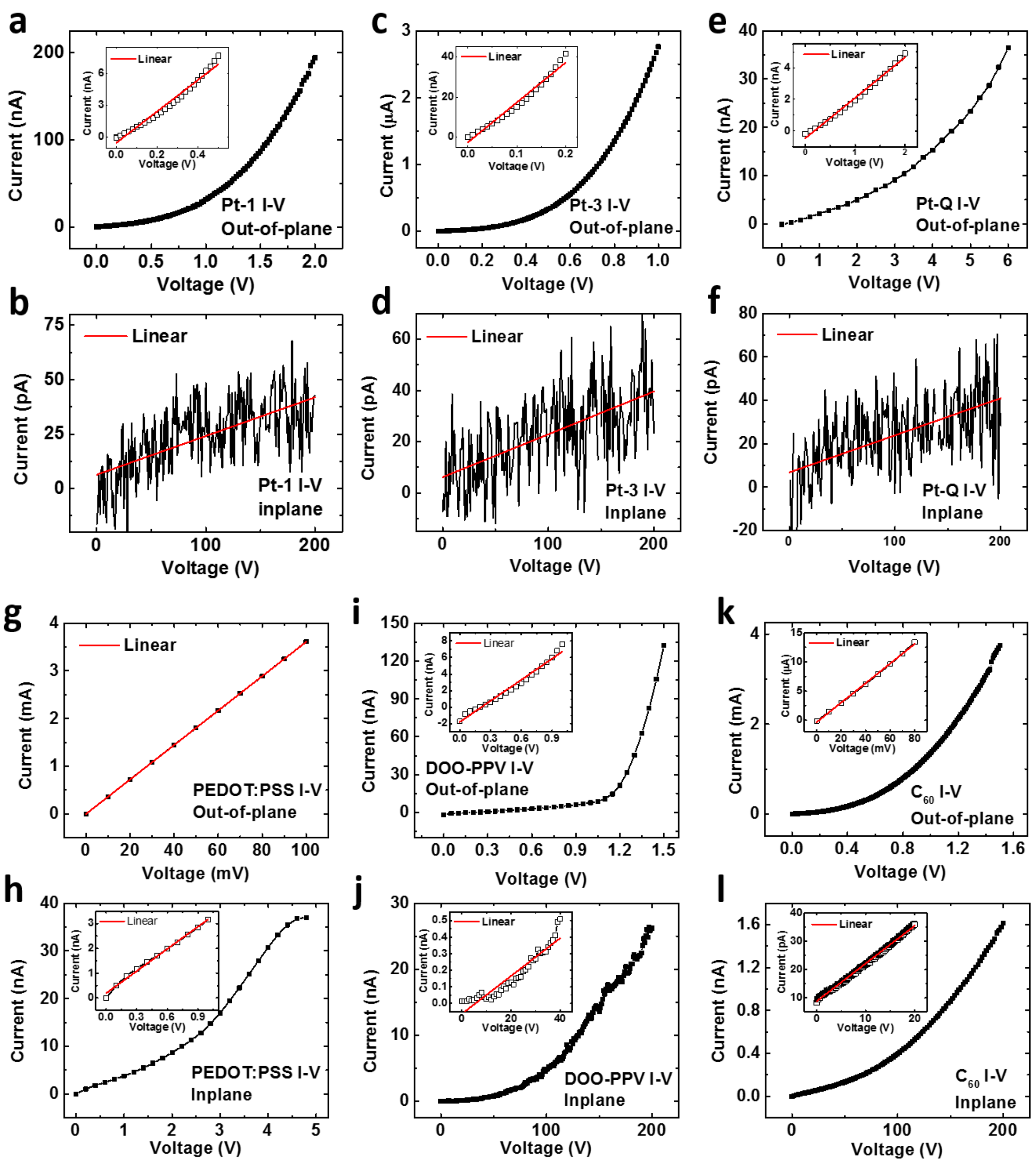

**Supplementary Figure S12 | *I-V* and conductivities for the OSEC materials. a, c, e, g, i, k,** *I-V* characteristic in OLED geometry for determining the out-of-plane conductivities in the OSEC materials. **b, d, f, h, j, l,** *I-V* characterization using four probe geometry for in-plane conductivities measurements in the OSEC materials. The insets show the *I-V* at low bias (open squares) with a linear fit (red line). The obtained conductivities are listed in Table S2.

## S6. Calculation of Spin Hall angle ($\theta_{SH}$) in Pt and the studied OSEC materials

| Materials | MW Frequency (GHz) | MW B1 (mT) | $4\pi M_{eff}$ (mT) | $\omega/\gamma$ (mT) | $\Delta B_{NiFe/X} - \Delta B_{NiFe}$ (mT) | $B_{res}$ (mT) | $\alpha$ | $\Delta 4\pi M_{eff}$ (mT) | $\Delta\theta_B$ (°) | $g_0$ | $J_S^0$ |
|---|---|---|---|---|---|---|---|---|---|---|---|
| Pt | 9.65 | 1.1 | 6369 | 316.229 | 1.0 | 116.0 | 0.0066 | 22 | 1.2 | $8.49\times10^{18}$ | $3.61\times10^{-7}$ |
| Pt-1 | 9.65 | 1.1 | 5586 | 316.232 | 3.3 | 115.1 | 0.0119 | 23 | 1.5 | $3.76\times10^{18}$ | $5.12\times10^{-7}$ |
| Pt-3 | 9.65 | 1.1 | 5703 | 316.229 | 2.0 | 117.4 | 0.0093 | 23 | 1.4 | $2.28\times10^{19}$ | $5.07\times10^{-7}$ |
| Pt-Q | 9.65 | 1.1 | 5479 | 316.232 | 1.7 | 126.0 | 0.0090 | 22 | 1.2 | $1.93\times10^{19}$ | $4.54\times10^{-7}$ |
| PEDOT:PSS | 9.65 | 1.1 | 5661 | 316.230 | 1.6 | 112.9 | 0.0088 | 23 | 1.8 | $1.81\times10^{19}$ | $4.47\times10^{-7}$ |
| DOO-PPV | 9.65 | 1.1 | 6200 | 316.239 | 6.2 | 140.6 | 0.0166 | 15 | 4.9 | $7.82\times10^{19}$ | $5.28\times10^{-7}$ |
| C60 | 9.65 | 1.1 | 5709 | 316.239 | 0.1 | 111.7 | 0.0010 | 24 | 2.0 | $5.80\times10^{17}$ | $1.11\times10^{-6}$ |

**Supplementary Table S1 | FMR response in OSEC p-ISHE devices and the derived parameters using Eqs. S7-S17.** The saturation magnetization ($4\pi M_S$), $\omega/\gamma$, broadened width between NiFe/OSCE and NiFe FMR spectra, FMR resonance fields are obtained from Fig. S8 (left panel) using Eqs. (S10) to (S14). The damping factor $\alpha$, magnetization deviation $\Delta 4\pi M_{eff}$ and angle $\Delta\theta_H$ are fitted from the data of the FMR width vs. $\theta_H$ (right panel in Fig. S10) using Eq. (S15) to (S17). The induced spin current density ($J_S^0$) and mixing conductance $g_r^{\uparrow\downarrow}$ at the NiFe/OSEC interface are calculated using Eqs. (S7) and (S8).

| Materials | SOC Ph/FL | $V_{ISHE}$ (μV) | ISHE-Is (nA) | $\omega C_N^{SUM} R_N^{SUM}$ (kHz*nF*Ω) | $R_N^{SUM}$ (Ω) | $\lambda_{(1)}$ (nm) | $\lambda_{(2)}$ (nm) | $\sigma_{in}$ (S m$^{-1}$) | $\sigma_{out}$ (S m$^{-1}$) | $\theta_{SH(1)}$ | $\theta^*_{SH(1)}$ | $\theta_{SH(2)}$ | $\theta^*_{SH(2)}$ |
|---|---|---|---|---|---|---|---|---|---|---|---|---|---|
| Pt | N/A | 1614 | 2892 | N/A | N/A | 2±0.5 | 3.4±0.4 | $2.0\times10^{6}$ | $2.0\times10^{6}$ | $+3.5\pm1.0\times10^{-2}$ | $+3.5\pm1.0\times10^{-2}$ | $+2.2\pm0.2\times10^{-2}$ | $+2.2\pm0.2\times10^{-2}$ |
| Pt-1 | 27 | 76 | 245 | $5.2\times10^{5}$ | 264 | 4±1 | 1.5±0.5 | $1.8\times10^{-8}$ | $2.1\times10^{-10}$ | $-4.4\pm0.9\times10^{-4}$ | $-5.1\pm1.1\times10^{-6}$ | $-1.2\pm0.3\times10^{-3}$ | $-1.4\pm0.4\times10^{-5}$ |
| Pt-3 | 12 | 52 | 231 | $2.7\times10^{5}$ | 161 | 5±1 | 2±0.5 | $6.0\times10^{-8}$ | $1.0\times10^{-9}$ | $-2.6\pm0.4\times10^{-4}$ | $-4.3\pm0.7\times10^{-6}$ | $-6.2\pm1.5\times10^{-4}$ | $-1.0\pm0.1\times10^{-5}$ |
| Pt-Q | 0.75 | 26 | 145 | $1.5\times10^{5}$ | 109 | 10±2 | 10±2 | $7.6\times10^{-9}$ | $8.3\times10^{-11}$ | $-7.1\pm1.5\times10^{-5}$ | $-7.7\pm1.5\times10^{-7}$ | $-7.1\pm1.3\times10^{-5}$ | $-7.7\pm1.5\times10^{-7}$ |
| PEDOT:PSS | N/A | 17 | 68 | $4.3\times10^{5}$ | 184 | 40±10 | 27±4 | $1.5\times10^{-3}$ | $5.6\times10^{-4}$ | $+2.2\pm0.2\times10^{-5}$ | $+8.2\pm0.8\times10^{-6}$ | $+2.4\pm0.2\times10^{-5}$ | $+9.1\pm0.6\times10^{-6}$ |
| DOO-PPV | N/A | 15 | 54 | $5.3\times10^{4}$ | 206 | 25±10 | 10±1 | $7.0\times10^{-7}$ | $8.0\times10^{-11}$ | $-1.4\pm0.4\times10^{-5}$ | $-1.5\pm0.5\times10^{-9}$ | $-3.3\pm0.3\times10^{-5}$ | $-3.8\pm0.4\times10^{-9}$ |
| C60 | N/A | 209 | 668 | $6.9\times10^{4}$ | 245 | 5±2 | 5±2 | $2.9\times10^{-7}$ | $5.0\times10^{-6}$ | $+4.5\pm1.5\times10^{-4}$ | $+7.7\pm1.5\times10^{-3}$ | $+4.5\pm1.5\times10^{-4}$ | $+7.7\pm1.5\times10^{-3}$ |

**Supplementary Table S2 | Spin Hall angle calculation in Pt and OSECs using the p-ISHE devices.** The relative SOC strengths in Pt-polymers are calculated from EL emission spectra presented in Fig. 3(a). $\omega_{SUM}$, $C_N^{SUM}$and $R_N^{SUM}$ are capacitance and resistance summations over the frequency range (100 Hz to 1 MHz) considering the weight of the p-ISHE intensity vs. frequency obtained from Figs. S5 and S6. The spin diffusion lengths $\lambda_{(1)}$ are fitted using Eq. (S6) from Fig. S8. $\lambda_{(2)}$ are the spin diffusing lengths estimated from OSVs and literatures reports. In-plane and out-of-plane conductivities are measured from Fig. S12 using OLED and four-probe geometries, respectively. The effective spin Hall angles $\theta_{SH(1)}$ and $\theta_{SH(2)}$ (that include the conductivity anisotropy) are calculated from Eq. (S6) using $\lambda_{(1)}$ and $\lambda_{(2)}$, respectively. $\theta^*_{SH(1)}$ and $\theta^*_{SH(2)}$ are the internal spin Hall angles after renormalization by the conductivity anisotropy between the in-plane and out-of-plane, where $\theta^*_{SH(1,2)} = \theta_{SH(1,2)} \frac{\sigma_{out}}{\sigma_{in}}$.